\documentclass[aps,prd,superscriptaddress,amsmath,amssymb,twocolumn,nofootinbib]{revtex4-1}
\usepackage{amsthm,graphicx,multirow}
\usepackage{epsfig}
\usepackage{latexsym, amssymb} 
\usepackage{amsmath}
\usepackage{epstopdf}
\usepackage[perpage]{footmisc}
\usepackage{xcolor}

\usepackage{hyperref}
\def\beq{\begin{equation}}
\def\eeq{\end{equation}}
\def\benu{\begin{enumerate}}
\def\efnu{\end{enumerate}}

\def\d{{\rm d}}

\newcommand{\dd}{\mathrm{d}}

\begin{document}
\renewcommand{\arraystretch}{1.4}

\title{Measuring lensing ratios with future cosmological surveys}

\author{Jos\'e Ram\'on Bermejo-Climent}\email{jose.bermejo@inaf.it}
\affiliation{INAF/OAS Bologna, Via Piero Gobetti 101, Area della Ricerca CNR/INAF, I-40129 Bologna, Italy}
\affiliation{INFN, Sezione di Bologna, Via Irnerio 46, I-40126 Bologna, Italy}
\affiliation{Departamento de Astrof\'isica, Universidad de La Laguna, 38206 La Laguna, Tenerife, Spain}
\author{Mario Ballardini}\email{mario.ballardini@inaf.it}
\affiliation{Dipartimento di Fisica e Astronomia, Alma Mater Studiorum Universit\'a di Bologna, Via Gobetti, 93/2, I-40129 Bologna, Italy}
\affiliation{INAF/OAS Bologna, Via Piero Gobetti 101, Area della Ricerca CNR/INAF, I-40129 Bologna, Italy}
\affiliation{INFN, Sezione di Bologna, Via Irnerio 46, I-40126 Bologna, Italy}
\affiliation{Department of Physics \& Astronomy, University of the Western Cape, Cape Town 7535, South Africa}
\author{Fabio Finelli}\email{fabio.finelli@inaf.it}
\affiliation{INAF/OAS Bologna, Via Piero Gobetti 101, Area della Ricerca CNR/INAF, I-40129 Bologna, Italy}
\affiliation{INFN, Sezione di Bologna, Via Irnerio 46, I-40126 Bologna, Italy}
\author{Vincenzo Fabrizio Cardone}\email{vincenzo.cardone@inaf.it}
\affiliation{INAF Osservatorio Astronomico di Roma, via Frascati 33, 00040 Monte Porzio Catone (Roma), Italy}

\begin{abstract}
The ratio between the CMB lensing/galaxy counts and the galaxy shear/galaxy counts cross-correlations combines the information from different cosmological probes to infer cosmographic measurements that are less dependent on astrophysical uncertainties and can constrain the geometry of the Universe. We discuss the future perspectives for the measurement of this lensing ratio as previously introduced, i.e. with the use of the Limber and flat-sky approximations and neglecting all the effects on the galaxy survey from observing on the past light cone. 
We then show how the cosmological information in this estimator is affected by the Limber approximation and by the inclusion of the redshift space distortions (RSD) and lensing magnification contributions to the galaxy number counts. We find that the lensing magnification contribution induces a multipole dependence of the lensing ratio that we show to be detectable at a statistical significant level combining post-$Planck$ CMB surveys and a Euclid-like experiment.
We propose an improved estimator which takes into account this angular scale dependence. Using this extended formalism, we present forecasts for upcoming and future cosmological surveys, and we show at which 
extent the lensing ratio information can improve the CMB constraints on cosmological parameters. We get that for extended cosmological models where the neutrino mass, the spatial curvature and the dark energy equation of state are allowed to vary, the constraints from $Planck$ on these parameters and on $H_0$ can be reduced by $\sim 40\%$ with the inclusion of a single lensing ratio and by $\sim 60$-70\% adding the joint measurement of 9 lensing ratios with a Euclid-like survey. We also find that neglecting the contribution from lensing magnification can induce a bias on the derived cosmological parameters in a combined analysis.
\end{abstract}

\pacs{Valid PACS appear here}% PACS, the Physics and Astronomy
                             % Classification Scheme.
\keywords{Suggested keywords}%Use showkeys class option if keyword
                              %display desired

\maketitle

\section{Introduction}

Weak gravitational lensing is one of the most direct probes of the distribution of dark matter
and it is correlated with the intervening process of structure formation.
Ratios between cross-correlations of galaxies and weak lensing at two different source planes in redshift have been proposed as cosmographic distance measurements \cite{Jain:2003tba,Bernstein:2003es,Zhang:2003ii,Bernstein:2005en,Hu:2007jh,Das:2008am}.
The role of these ratio estimators between the weak lensing at two different redshift and
a matter tracer as a cosmographic measure 
becomes extremely transparent under different approximations, such as the Limber and
flat sky approximation, and the limit in which the foreground distribution is extremely peaked in redshift.

Being a ratio between two cross-correlation terms with the same
lens, this estimator is largely independent
on the clustering bias of the lens and weak lensing systematics, but depends on most of the background cosmological parameters.
By taking one of the source planes as the CMB last scattering surface, the lever arm of such a lensing ratio estimator becomes somewhat maximal \cite{Hu:2007jh,Das:2008am}.

The scientific potential of the CMB lensing ratio as a cosmographic measurement for the
next generation of CMB and LSS experiments has been forecast in several papers
\cite{Das:2008am,Ade:2018sbj,Prat:2018yru}. %In the meantime, 
The estimator for the lensing ratio between CMB lensing/galaxies and galaxy shear/galaxies
has already been applied to real data \cite{Miyatake:2016gdc,Prat:2018yru}. 
Miyatake et al. \cite{Miyatake:2016gdc} used 
CMASS \cite{1010.4915} and CFHTLens \cite{1210.8156} for galaxy lenses and sources, respectively, and CMB lensing from $Planck$ 2015 \cite{1502.01591}. Prat et al. \cite{Prat:2018yru} used galaxy position and lensing from DES Y1 \cite{1708.01530} and CMB lensing from a combination of $Planck$ 2015 and SPT \cite{1705.00743}.

In this paper we study and extend the lensing ratio estimator as introduced by Das and Spergel in \cite{Das:2008am} (henceforth DS) and we study its scientific capabilities in the context of future cosmological observations.
We show how the approximations in the galaxy and lensing kernel and the finite width in redshift of the lenses
density distributions affect to the multipole dependence of the lensing ratio.
The inclusion of the lensing magnification contribution
in the galaxy number counts introduces a further and larger dependence on the multipoles, which we show to be detected with future cosmological observations, and calls for an extension of a lensing ratio estimator which
takes into account the dependence on multipoles.
%Since relativistic corrections will be detectable in future surveys

Our paper is organized as follows. After this introduction, we introduce the notation for the CMB lensing/galaxy
and galaxy shear/galaxy cross-correlation, respectively, in Section \ref{sec:formalism}. In Section \ref{sec:data} we introduce 
the experimental specifications of the CMB anisotropies and galaxy surveys we use in our forecasts. 
In Section \ref{sec:DS} we forecast the capabilities of a Euclid-like\footnote{\href{https://www.cosmos.esa.int/web/euclid/home}{https://www.cosmos.esa.int/web/euclid/home}} \cite{1110.3193} experiment alone and in combination with
galaxy lenses at lower redshift from DESI\footnote{\href{https://www.desi.lbl.gov/}{https://www.desi.lbl.gov/}} \cite{1611.00036} and SPHEREx\footnote{\href{http://spherex.caltech.edu/}{http://spherex.caltech.edu/}} \cite{1412.4872} in measuring the lensing ratio as originally introduced in DS.
In Section \ref{sec:generalized} we consider the ratio between the CMB lensing/galaxy
and galaxy shear/galaxy cross-correlations without approximations and replacing the galaxy density
with the galaxy number counts including RSD and lensing magnification contributions and introduce its optimal estimator and minimum variance.
In Section \ref{sec:cosmoforecast} we forecast the expected errors on cosmological parameters by using the novel methodology
introduced in Section \ref{sec:generalized}. In Section \ref{sec:conclusions} we draw our conclusions.

\section{Formalism}
\label{sec:formalism}
In this section we define the quantities involved in the angular power spectra of the cross-correlation between a foreground lens galaxy population and a background weak lensing source that comes from the CMB or from the galaxy shear.
We define the angular power spectrum as
\begin{equation}
    \delta_{\ell \ell'}\delta_{mm'} C_\ell^a = \langle a_{\ell m}a^*_{\ell' m'}\rangle
\end{equation}
where $a_{\ell m}$ are the spherical harmonic coefficients obtained from the expansion of 
a scalar field with spin-0 spherical harmonics as
\begin{equation}
    a\left(\hat{\bf n}\right) = \sum_{\ell m} a_{\ell m} Y^*_{\ell m}\left(\hat{\bf n}\right) \,.
\end{equation}

We are interested in the cross-correlation of a foreground galaxy number density field with 
two different backgrounds as the convergence field from the weak lensing of galaxies and of the CMB. 
The angular power spectrum can be calculated as
\begin{equation} \label{eqn:APS}
    C_\ell^{XY} = 4\pi \int \frac{\dd k}{k} {\cal P}(k) I^X_{\ell}(k) I^Y_{\ell}(k)
\end{equation}
where ${\cal P}(k) \equiv k^3 P(k) / (2\pi^2)$ is the dimensionless primordial power spectrum 
and $I^X_{\ell}(k)$ is the kernel for the $X$ field for unit primordial power spectrum.

All the weak lensing quantities can be defined from the lensing potential
\begin{equation}
    \phi\left(\hat{\bf n},\chi\right) = \frac{2}{c^2} \int_0^\chi \dd \chi' \frac{\chi-\chi'}{\chi\chi'}
    \Phi\left(\chi'\hat{\bf n},\chi'\right)
\end{equation}
where $\Phi\left(\hat{\bf n},\chi\right)$ is the gravitational potential.
% at comoving position ${\bf x}$ and conformal time $\chi$. 
The comoving distance is
\begin{equation}
    \chi(z) = \int_0^z \frac{c\, \dd z'}{H(z')}\,. 
\end{equation}

The observable 2-dimensional lensing potential, averaged over background sources with a redshift 
distribution $W_b\left(\chi\right)$, 
is given by
\begin{equation}
    \phi\left(\hat{\bf n}\right) = \frac{2}{c^2} \int_0^\chi \frac{\dd \chi'}{\chi'} q_b\left(\chi'\right)
    \Phi\left(\chi'\hat{\bf n},\chi'\right)
\end{equation}
where $q_b\left(\chi\right)$ is the lensing efficiency (for a given background distribution $W_b$) 
defined as
\begin{equation}
    q_b\left(\chi\right) = \int_\chi \dd \chi' \frac{\chi'-\chi}{\chi'} W_b\left(\chi'\right) \,.
\end{equation}

By expanding the gravitational potential in Fourier space and using the plane-wave expansion, we can 
define the lensing potential kernel as \cite{Jain:1996st}
\begin{equation} \label{eqn:kernel_phi}
    I^\phi_\ell (k) = 2\left(\frac{3\Omega_mH_0^2}{2k^2c^2}\right) \int \frac{\dd \chi}{(2\pi)^{3/2}}
    \frac{q_b\left(\chi\right)}{\chi a(\chi)} j_\ell\left(k\chi\right) 
    \delta\left(k,\chi\right) \,,
\end{equation}
where $\Omega_m$ is the present-day matter density, $H_0$ is the Hubble constant, $\delta(k,\chi)$ 
is the comoving-gauge matter density perturbation, and $j_\ell$ the spherical Bessel functions. 
In case of CMB lensing, the source distribution can be approximated by 
$W_{\rm CMB} \left(\chi\right) \simeq \delta_{\rm D}\left(\chi-\chi_*\right)$ and the lensing efficiency by
\begin{equation}
    q_{\rm CMB}\left(\chi\right) \simeq \frac{\chi_*-\chi}{\chi_*}
\end{equation}
where $\chi_*$ is the comoving distance at the surface of last scattering, and Eq.~\eqref{eqn:kernel_phi} 
reduces to
\begin{multline}
 \label{eqn:kernel_CMB}
    I^{\phi_{\rm CMB}}_\ell (k) = 2\left(\frac{3\Omega_mH_0^2}{2k^2c^2}\right) \\ \times \int \frac{\dd \chi}{(2\pi)^{3/2}} \frac{\chi_*-\chi}{\chi_*\chi} \frac{1}{a(\chi)}
    j_\ell\left(k\chi\right) \delta\left(k,\chi\right) \,.
\end{multline}
Finally the convergence $\kappa = \nabla^2 \phi/2$ can be expanded in spherical-harmonics as
\begin{equation}
    \kappa\left({\bf \hat{n}}\right) = -\frac{1}{2} \sum_{\ell, m} \ell(\ell+1)\phi_{\ell m}Y_\ell^{m}\left({\bf \hat{n}}\right)
\end{equation}
and we can relate the two kernel functions by
\begin{equation} \label{eqn:kernel_kappa}
    I_\ell^{\kappa}(k) = \frac{\ell(\ell+1)}{2} I_\ell^{\phi}(k) \,.
\end{equation}

The 2-dimensional integrate window function for the galaxy number counts is
\begin{equation} \label{eqn:kernel_counts}
    I^G_\ell(k) = \int \frac{\dd \chi}{(2\pi)^{3/2}} W_f(\chi) \Delta^s_{\ell}(k,\chi) 
\end{equation}
where $\Delta^s_{\ell}(k,\chi)$ is the synchronous gauge source counts Fourier transformed and expanded into multipoles and $W_f(\chi)$ is the foreground redshift distribution of galaxies. We assume that $\Delta^s_{\ell}(k,\chi)$ is related to the underlying matter density field through a redshift dependent galaxy bias $b_g$ as $\Delta^s_{\ell}(k,\chi) = b_g(\chi) \delta(k,\chi) j_\ell\left(k\chi\right)$.

Finally, we define the lensing ratio as DS
\begin{equation}
    \label{eq:lensingratio}
    r_\ell \equiv \frac{C_\ell^{\kappa_{\rm CMB} G}}{C_\ell^{\kappa_{\rm gal} G}} \,.
\end{equation}

\section{Data and specifications}
\label{sec:data}
We define here the specifications for the future large scale structure and CMB surveys considered in order to produce 
the mock signal and noise data. The lensing ratio estimator is based in the cross-correlations between three ingredients: 
a tracer for the foreground galaxy population, a background of source galaxies traced by a Euclid-like photometric survey; 
and the CMB lensing background source, for which we consider a $Planck$-like experiment and many future experiments.

We create the mock data for the angular power spectra using \texttt{CLASSgal} \cite{Blas:2011rf,1307.1459}. 
The non-linear corrections are modeled as \texttt{halofit} with the recipe by \cite{1208.2701}. For the fiducial 
cosmology we assume a $\Lambda$CDM+$\sum m_\nu$ model with one massive neutrino consistent with the $Planck$ 2018 results \cite{1807.06209}. We use
$\Omega_b h^2$ = 0.022383, $\Omega_c h^2$ = 0.12011, $H_0$ = 67.32 km s$^{-1}$ Mpc$^{-1}$, $\tau$ = 0.0543, 
$n_s$ = 0.96605, $\ln (10^{10} A_s)$ = 3.0448 and $\sum m_\nu$ = 0.06 eV.

\subsection{Galaxy lenses}
For the foreground lens population we use a Euclid-like spectroscopic survey and lower redshift populations like DESI and SPHEREx that allow to increase the background number of objects and the distance between the lens and the sources, which come from the Euclid photometric survey. We describe here the specifications of these experiments.

We adopt as baseline for a given lens population narrow slices with $\Delta z$ = 0.1. For this, we convolve the number density distribution $\d N/\d z$ with a Gaussian probability distribution for the measured redshift given the redshift accuracy. Following \cite{Ma:2005rc}, the number density distribution of a single bin is expressed as

\begin{equation}
\label{eq:binning}
\frac{\d n_{\rm gal}^i}{\d z} = \frac{\d N}{\d z} \int_{z_{\rm min}}^{z_{\rm max}}\d z_m p(z_m|z) \,,
\end{equation}
where $z_{\rm min}$, $z_{\rm max}$ are the edges of the redshift bin and $p(z_m|z)$ is the probability density for the measured redshift $z_m$ given the true redshift $z$ of the galaxy, given by
\begin{equation}
\label{eq:photo-z}
p(z_m|z) = \frac{1}{\sqrt{2 \pi} \sigma_z} e^{-\frac{1}{2}(z_m - z)^2 / \sigma_z^2}\,.
\end{equation}
Inserting Eq.~\eqref{eq:photo-z} into Eq.~\eqref{eq:binning} we obtain:

 \begin{equation}
\label{eq:densitybins}
\frac{\d n_i}{\d z} = \frac{1}{2} \frac{\d N}{\d z} \left[ {\rm erf} \left( \frac{z_{\rm max} - z}{\sqrt{2} \sigma_z } \right) - {\rm erf} \left( \frac{z_{\rm min} - z}{\sqrt{2} \sigma_z } \right) \right]
  \end{equation} 
where erf is the error function.

In the harmonic space, the Poisson shot noise for a given foreground population at redshift $z_i$ is obtained as the inverse of the number of objects per steradian,
 \begin{equation}
{\cal N}_{\ell}^{G}(z_i)  = \frac{4 \pi f_{\rm sky}}{\bar{n}_{\rm g}^{i}} 
\end{equation}
 where $f_{\rm sky}$ is the sky fraction and ${n_{\rm g}^{i}}$ is the total number of galaxies.
\subsubsection{Euclid-like spectroscopic survey}
\label{sec:euclid_sp}

The Euclid spectroscopic survey will measure the galaxy clustering from millions of $H\alpha$ emitters in a redshift range $0.9 \le z \le 1.8$ with a sky coverage of 15000 deg$^2$. The number density distribution  $\d N/\d z$ of the survey is fitted from the {\em model 3} data by \cite{Pozzetti:2016cch} using a flux threshold $F_{\textup{H}\alpha} > 2 \times 10^{-16}\ \text{erg}\ \text{cm}^{-2}\ \text{s}^{-1}$. This yields as total number density of objects $\bar{n}_\mathrm{g} = 2039$ sources per deg$^2$, for which we introduce a 50\% factor due to the Euclid completeness and purity. We assume a bias evolution function $b_g(z)= 0.7 + 0.7z$  according to the fitting for H$\alpha$ emission line object from \cite{1903.02030}. The redshift accuracy is characterized by a dispersion $\sigma_z = 0.001 (1+z)$. We represent in Fig.~\ref{fig:configuration} the $\d N/\d z$ of the full survey and the selected foreground population for the first redshift bin at 0.9 $< z_{\rm lens} <$ 1. We will refer hereafter this foreground configuration as Euclid-r1.

 \subsubsection{Lenses at lower redshift}
For the foreground lens populations at lower redshift we consider the ground-based survey Dark Energy Spectroscopic Instrument (DESI) and the recently approved NASA mission Spectro-Photometer for the History of the Universe, Epoch of Reionization, and Ices Explorer (SPHEREx).

DESI is an ongoing spectroscopic survey that covers $\sim 14000$ deg$^2$ in the sky. Here we consider the lower redshift target objects: the Bright Galaxy Sample (BGS), which will measure $\sim 10$ million galaxies at 0 $< z <$ 0.4. We adopt the specifications in \cite{1611.00036} for the number density distribution and the  bias redshift evolution, which is given by $b_g(z) = 1.34/D(z)$. The redshift accuracy is given by $\sigma_z = 0.001 (1+z)$. The overlapping sky fraction with Euclid will be limited to $\sim 4000$ deg$^2$.

SPHEREx will be a full-sky spectro-photometric survey that can operate with different configurations depending on the number of objects and redshift accuracy. In this work we assume SPHEREx-2, a configuration with $\sigma_z = 0.008 (1+z)$ and $\sim 70$ million objects\footnote{For the SPHEREx number density distribution and the bias redshift evolution we fit the data by Olivier Dor\'e (private communication).}. Since this survey will cover $\sim 80\%$ of the sky, there will be full overlap with the background from Euclid and all the CMB experiments. We represent in Fig.~\ref{fig:configuration} the $\d N/\d z$ of the full survey and the lens foreground population for a bin at 0.2 $< z_{\rm lens} <$ 0.3. We will refer hereafter this foreground configuration as SPHEREx-r1.

\subsection{Galaxy shear sources: Euclid-like photometric survey}

\label{sec:euclid_ph}
The Euclid photometric survey will measure both galaxy clustering and weak lensing from a sample of billions of galaxies. Here we will consider the weak lensing from a given background population. We parametrize the $\d N/\d z$ of the survey as
\begin{equation}
\label{eq:dNdz}
\frac{\d N}{\d z} \propto z^{\alpha} \exp \left[ -\left( \frac{z}{z_0} \right)^{\beta} \right]
\end{equation}
where $\alpha$ = 2, $\beta$ = 3/2, $z_0 = z_{\rm mean}/\sqrt[]{2}$ and $z_{\rm mean}$ = 0.9 is the mean redshift of the survey. The number density of the sources population is $\bar{n}_\mathrm{g}=30$ objects per arcmin$^2$. The sky coverage is 15000 deg$^2$ as well as for the spectroscopic survey, and the bias is assumed to evolve with redshift following $b_G(z)=\sqrt{1+z}$ \cite{1206.1225}. 

\begin{figure}
    \centering
    \includegraphics[width = \columnwidth]{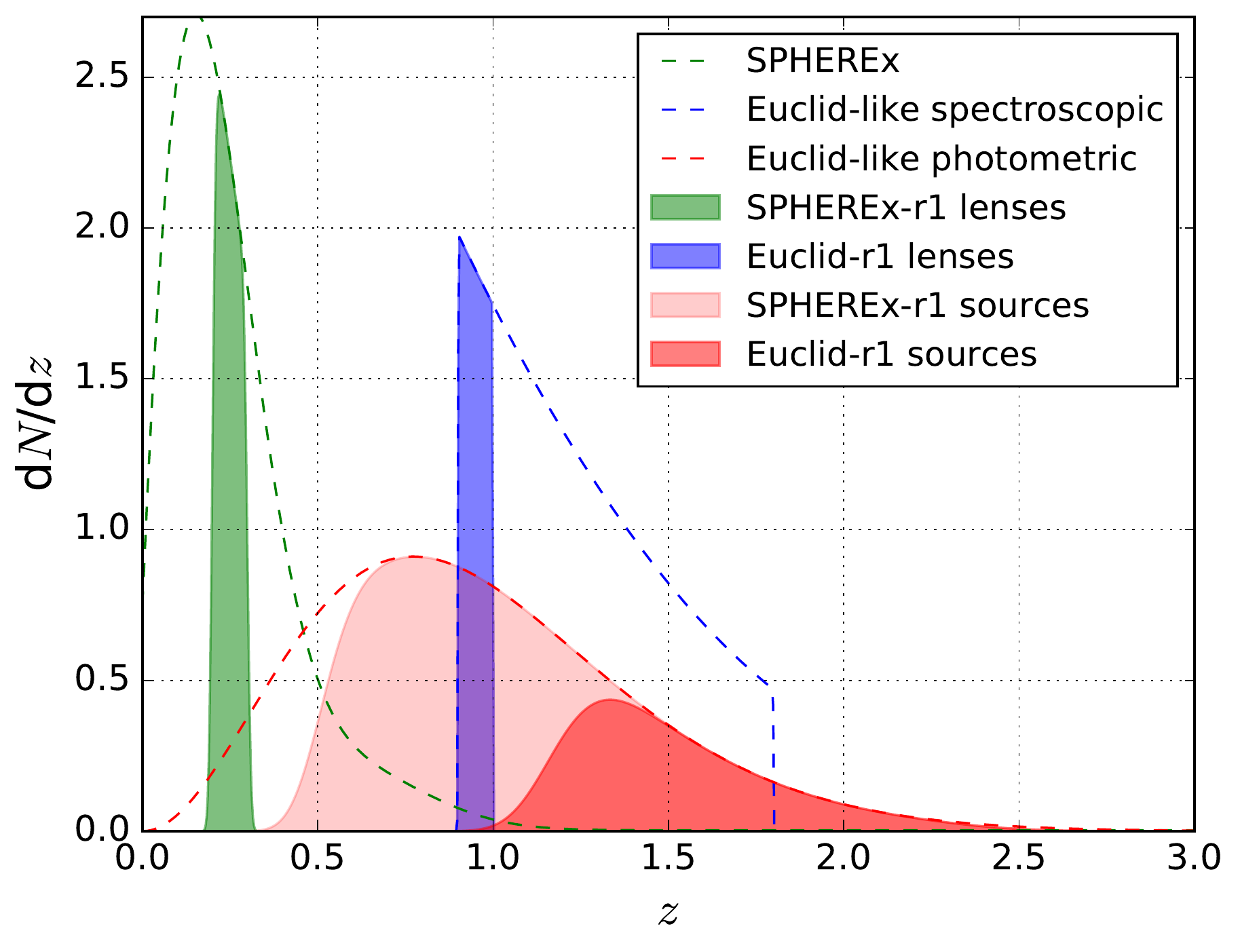}
    \caption{Survey configuration for the lensing ratio. The green dashed curve represents the normalized $\d N/\d z$ of SPHEREx and the green shaded area corresponds to a bin at 0.2 $< z <$ 0.3 for the foreground population of the SPHEREx-r1 configuration. The blue dashed curve represents the normalized $\d N/\d z$ of the Euclid-like spectroscopic survey, and the blue shaded area corresponds to the bin at 0.9 $< z <$ 1.0 that traces the foreground of the Euclid-r1 configuration. The red dashed curve represents the normalized $\d N/\d z$ of the Euclid-like photometric survey, and the pink and red shaded areas correspond to the background of source galaxies beyond the SPHEREx-r1 and Euclid-r1 lenses.}
    \label{fig:configuration}
\end{figure}

We assume that the background population is given by a broad bin that maximizes the number of objects behind the lenses without overlapping with them. For this, we convolve the $\d N/\d z$ of the photometric survey with a Gaussian redshift probability distribution with a dispersion $\sigma_z = 0.05 (1+z)$, following Eq.~\eqref{eq:densitybins}. We show in Fig.~\ref{fig:configuration} the $\d N/\d z$ of the photometric survey and the maximal background source populations for the two foregrounds of the Euclid-r1 and SPHEREx-r1 configurations.

The shear noise for the background population for a redshift bin at $z_i$ is obtained as
 \begin{equation}
{\cal N}_{\ell}^{\kappa_{\rm gal}}(z_i)  = \sigma_{\epsilon}^2 \frac{4 \pi f_{\rm sky}}{\bar{n}_{\rm g}^{i}} 
\end{equation}
 where $\sigma_\epsilon$ is the intrinsic ellipticity RMS, for which we adopt $\sigma_\epsilon$ = 0.22, $f_{\rm sky}$ is the sky coverage and ${n_{\rm g}^{i}}$ is the number of sources.

\subsection{CMB lensing source}

For the CMB lensing background source we consider as surveys the ESA mission $Planck$\footnote{\href{https://www.cosmos.esa.int/web/planck}{https://www.cosmos.esa.int/web/planck}} \cite{1807.06205}, the ground-based future experiments Simons Observatory (SO)\footnote{\href{https://simonsobservatory.org/}{https://simonsobservatory.org/}} \cite{1808.07445} and CMB Stage-4 (S4)\footnote{\href{https://cmb-s4.org/}{https://cmb-s4.org/}} \cite{1610.02743}, the proposed space missions Lite satellite for the studies of B-mode polarization and Inflation from cosmic background Radiation Detection (LiteBIRD)\footnote{\href{http://litebird.jp/eng/}{http://litebird.jp/eng/}} \cite{Matsumura:2016sri}, and the two concepts Probe of Inflation and Cosmic Origins (PICO) \footnote{\href{http://pico.umn.edu}{http://pico.umn.edu}}\cite{1902.10541} and Polarized Radiation Imaging and Spectroscopy Mission (PRISM)\footnote{\href{https://www.cosmos.esa.int/web/voyage-2050}{https://www.cosmos.esa.int/web/voyage-2050}} \cite{1909.01591}. The specifications of these experiments will be also used for temperature and polarization for a quantitative assessment of what the lensing ratio can add to the information of the CMB fields alone.

We reconstruct the minimum variance (MV) estimator for the CMB lensing noise ${\cal N}_{\ell}^{\phi\phi}$ using the temperature and polarization noise ${\cal N}_{\ell}^{TT}$ and ${\cal N}_{\ell}^{EE}$. This is done combining the $TT$, $EE$, $BB$, $TE$, $TB$, $EB$ estimators following the Hu-Okamoto algorithm \cite{astro-ph/0301031} and using the public code \texttt{quicklens}\footnote{\href{https://github.com/dhanson/quicklens}{https://github.com/dhanson/quicklens}}. For the $TT$ and $EE$ channels we calculate the isotropic noise deconvolved with the instrument beam using the formula
\begin{equation}
{\cal N}_{\ell}^X = w_X^{-1} b_{\ell}^{-2}, \qquad b_{\ell} = e^{-\ell(\ell+1)\theta_{\rm FWHM}^2/16 \ln 2}
\end{equation}
where $\theta_{\rm FWHM}$ is the FWHM of the beam in radians and $w_{TT}$, $w_{EE}$ are the inverse square of the detector noise level for temperature and polarization in arcmin$^{-1}$ $\mu$K$^{-1}$.

In order to provide specifications of simulated $Planck$-like data leading to uncertainties on the cosmological parameters compatible with the latest results in \cite{1807.06209} we adopt $w_{TT}$ = 33 $\mu$K arcmin, $w_{EE}$ = 70.2 $\mu$K arcmin and $\theta_{\rm FWHM} = 7.3$ arcmin for the 143 GHz channel and assume a sky fraction of $f_{\rm sky} = 0.7$. We re-adapt the noise in polarization ${\cal N}_{\ell}^{EE}$ inflating the noise in polarization at $\ell < 30$ in order to obtain estimates for the uncertainty in the optical depth compatible with the $Planck$ 2018 results.
%We consider the CMB lensing power spectrum ${C}_{\ell}^{\phi \phi}$ in the conservative range, i.e. for 
%$8 \le \ell \le 400$, and neglect the $T \phi$, $E\phi$ cross-correlations according to the $Planck$ real likelihood.
For the $Planck$-like survey, the effective noise bias for the CMB lensing power spectrum is obtained from the inverse weighted sum of the specifications of the 143 and 217 GHz channels in \cite{Adam:2015rua} to match the $Planck$ 2018 performances 
\cite{1807.06209}. We assume a sky coverage $f_{\rm sky}$ = 0.7 also for the lensing, which fully overlap with Euclid and we set $\ell_{\rm max}$ = 1500.

%For SO we adopt the specifications of the six frequency bands listed in \cite{1808.07445}. 
For SO we adopt the specifications of the six frequency bands listed in \cite{1808.07445}. We assume $f_{\rm sky}$ = 0.4 and $\ell_{\rm max}$ = 3000. Since this is a ground-based experiment, the largest scales will not be seen by SO, hence we set $\ell_{\rm min} = 30$ and consider the $Planck$ specifications for $2 \leq \ell \leq 29$ (hereafter we call $Planck$+SO to this combination). We rescale the MV noise bias for the CMB lensing to match the baseline configuration in 
\cite{1808.07445}. Given that the experiment will be based in the southern hemisphere, there will not be full overlap with the Euclid sky coverage. We assume that the common sky fraction will be around 25\% of the sky. 

For S4 we adopt $w_{TT}$ = 1 $\mu$K arcmin, $w_{EE} = \sqrt{2}$ $\mu$K arcmin, $\theta_{\rm FWHM} = 3$ arcmin \cite{1610.02743} and we assume as for SO $f_{\rm sky}$ = 0.4 and $\ell_{\rm max}$ = 3000. Since this experiment will be also based in the southern hemisphere, we limit as well the overlapping sky fraction with Euclid to 25\% and adopt the $Planck$ noise for $\ell <$ 30.

For LiteBird we combine the 7 channels described in \cite{1612.08270}. We assume 70\% sky fraction and since this mission will be optimized for large scales, we adopt $\ell_{\rm max}$ = 1350. 

For PICO we use the 7 channels ranging from 75 to 220 GHz given in \cite{1902.10541}. We assume $\ell_{\rm max}$ = 3000 and 70\% of sky coverage.

For PRISM we sum the 12 channels ranging from 52 to 385 GHz in \cite{1909.01591}. We adopt $\ell_{\rm max}$ = 4000 and 75\% for the sky fraction.

\begin{figure}
    \centering
    \includegraphics[width = \columnwidth]{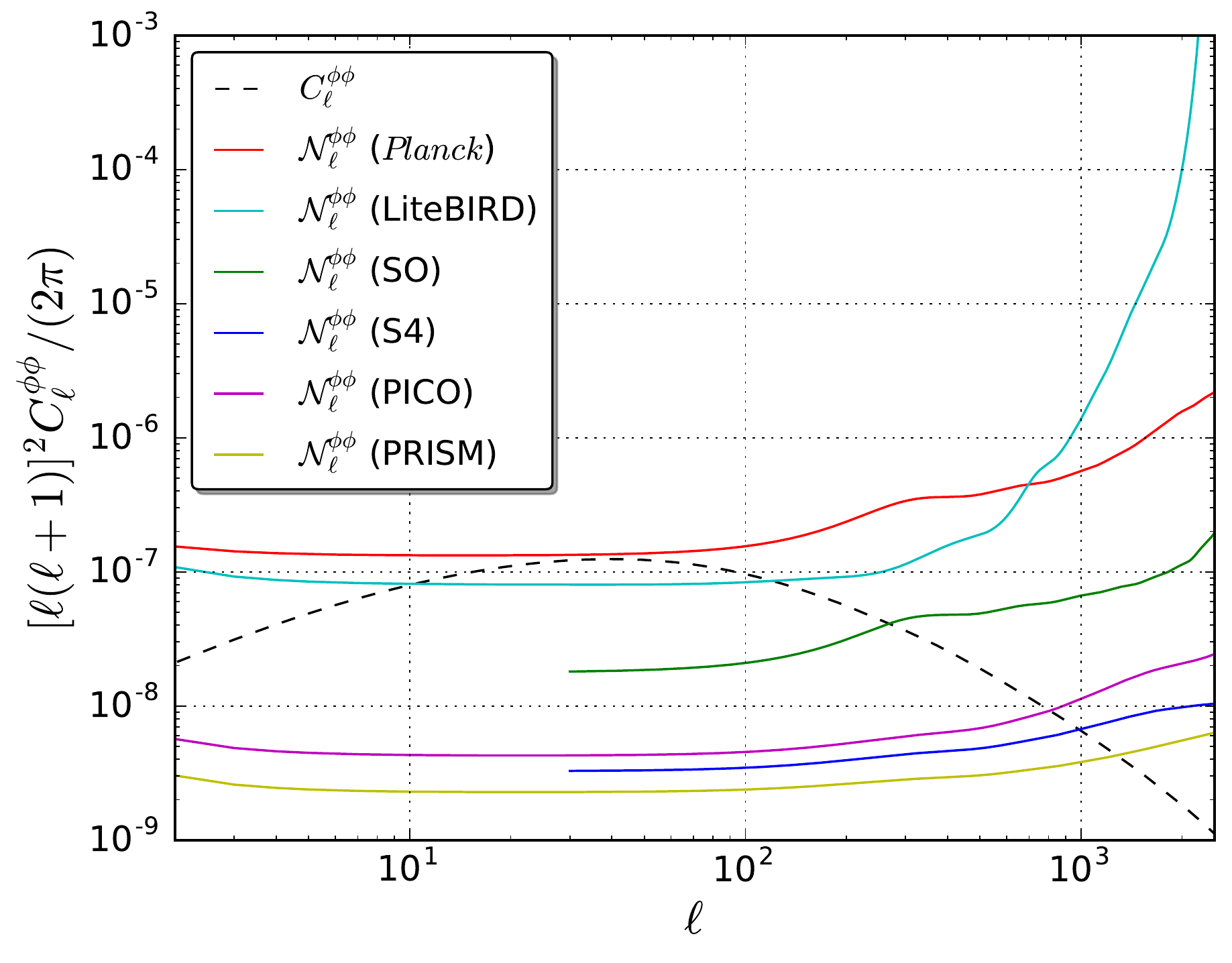}
    \caption{Signal of the CMB lensing potential data and its noise computed for the experiments considered using the minimum variance estimator.}
    \label{fig:cmb}
\end{figure}

We show in Fig.~\ref{fig:cmb} the CMB lensing potential noise ${\cal N}_{\ell}^{\phi\phi}$ obtained for the experiments described above.

\section{Cosmographic lensing ratio measurements}
\label{sec:DS}
In this section we study how under some approximations the lensing ratio $r_\ell$ can be interpreted as a cosmographic measurement that does not depend on the multipoles, astrophysical uncertainties and perturbations. We then present forecasts for the error on the lensing ratio for this previously introduced limit using the future cosmological surveys mock data described in Section~\ref{sec:data} and explore how this uncertainty varies with the foreground population redshift $z_{\rm lens}$ and the selected background.  

 \subsection{The cosmographic ratio limit}
 We show here the limit in which the lensing ratio $r_\ell$ defined in Section~\ref{sec:formalism} becomes a geometrical quantity independent of the angular scale, the power spectrum and the galaxy bias. This limit needs to assume the Limber approximation, to select a foreground lens population which is narrow enough in redshift and to neglect the effects on the galaxy number counts from observing 
on the past light cone.
 
\subsubsection{Limber approximation}

In order to speed up the computation of Eq.~\eqref{eqn:APS},  which is time-consuming due to the rapid oscillations 
of the spherical Bessel function at high multipoles, it is commonly adopted the Limber approximation 
\cite{Limber:1954zz} which is accurate at high-$\ell$. It consists in replacing the spherical Bessel function 
$j_\ell(k\chi)$ with a Dirac delta-function $\delta_{\rm D}$
\begin{equation} \label{eqn:Limber}
    j_\ell(k\chi) \to \sqrt{\frac{\pi}{2(\ell+1/2)}}\delta_{\rm D}\left(\ell+\frac{1}{2}-k\chi\right) \,.
\end{equation}
We can then approximate the kernel functions \eqref{eqn:kernel_phi}-\eqref{eqn:kernel_CMB}-\eqref{eqn:kernel_counts} 
obtaining the following angular power spectra
\begin{multline}
    C_\ell^{\phi \phi}(z_i,z_j) = \frac{4}{(\ell+1/2)^4} \left(\frac{3\Omega_{\rm m}H_0^2}{2c^2}\right)^2 \\ \times
        \int \dd \chi \frac{q_{b_i}(\chi)q_{b_j}(\chi)}{a^2(\chi)} 
        P_\delta\left(\frac{\ell+1/2}{\chi},\chi\right) \,,
\end{multline}
\begin{multline}
    C_\ell^{GG}(z_i,z_j) =\\ \int \dd \chi \frac{W_{f_i}(\chi)W_{f_j}(\chi)}{\chi^2} b_g^2(\chi)
        P_\delta\left(\frac{\ell+1/2}{\chi},\chi\right)\,,
\end{multline}
\begin{multline}
       C_\ell^{\phi G}(z_i,z_j) = \frac{2}{(\ell+1/2)^2} \left(\frac{3\Omega_{\rm m}H_0^2}{2c^2}\right) \\ \times
        \int \dd \chi \frac{q_{b_i}(\chi)W_{f_j}(\chi)}{a(\chi)\chi} b_g(\chi)
        P_\delta\left(\frac{\ell+1/2}{\chi},\chi\right) \,,
    \label{eq:limberphiG} 
\end{multline}

for background sources, foreground lenses, and their cross-correlation, where the matter power spectrum is defined as
\begin{equation}
    \langle \delta({\bf k},\chi)\delta^*({\bf k'},\chi) \rangle = 
        (2\pi)^3 P_\delta(k,\chi) \delta_{\rm D}({\bf k}-{\bf k'}) \,.
\end{equation}

\begin{figure*}
    \centering
    \includegraphics[width = \textwidth]{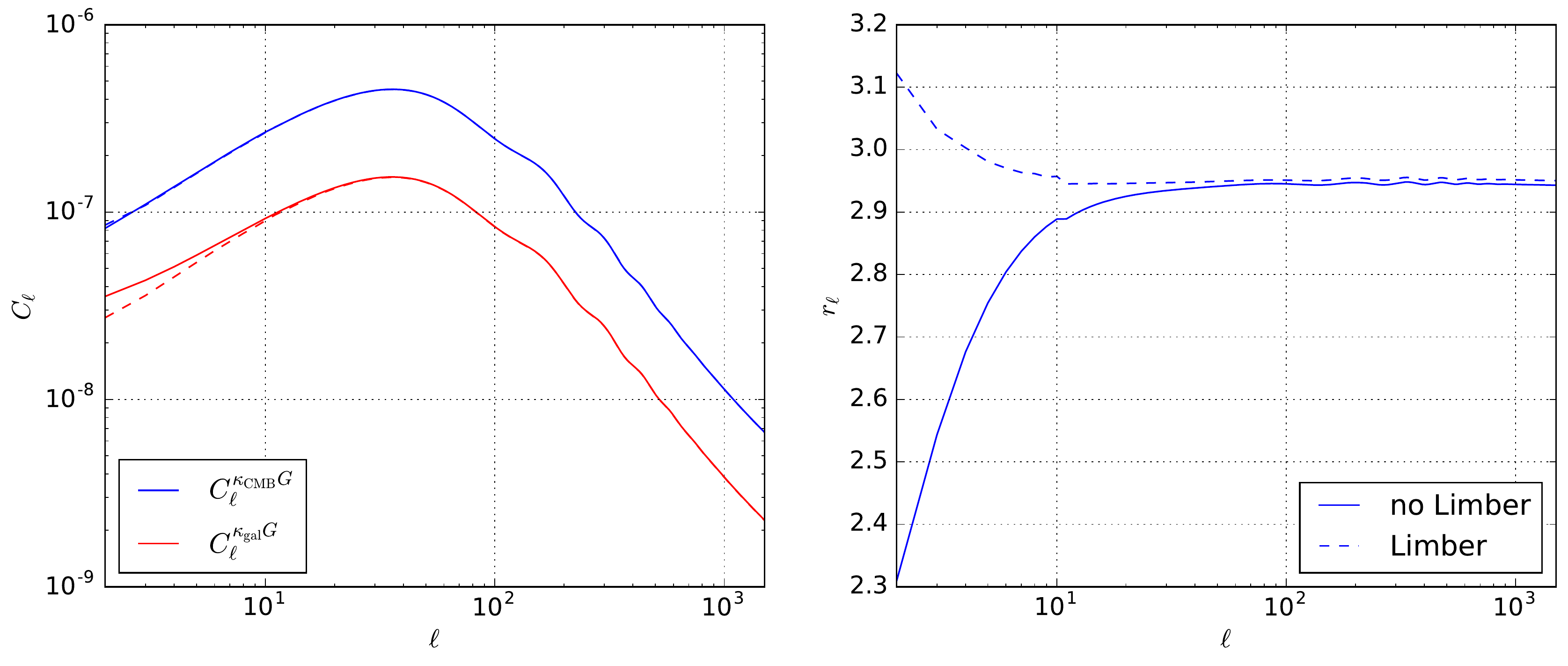}
    \caption{Impact of the Limber approximation on both lensing-galaxy cross-correlation angular power spectra (left panel) and on the lensing ratio (right panel), using the Euclid-r1 configuration at $z_{\rm lens}$ = 0.95 }
    \label{fig:limber}
\end{figure*}

We show in Fig.~\ref{fig:limber} the effect of the Limber approximation in the cross-correlation angular power spectra $C_\ell^{\kappa_{\rm CMB} G}$ and $C_\ell^{\kappa_{\rm gal} G}$ and in the lensing ratio $r_\ell$, using the Euclid-r1 configuration. We find that the Limber approximation changes the signal of the denominator $C_\ell^{\kappa_{\rm gal} G}$ and hence the ratio $r_\ell$ at the lowest multipoles, smoothing the $\ell$-dependence that appears when this approximation is not used.

In DS it is also considered the flat-sky approximation. In this limit, the sky is approximated by a 2-dimensional plane tangential to the celestial sphere 
and mathematically expansions in spherical harmonics are replaced by Fourier expansions
\begin{equation}
    \sum_{\ell,m} \phi_{\ell,m}Y_\ell^m(\hat{\bf n}) \to 
    \int \frac{\dd^2 {\boldsymbol \theta}}{(2\pi)^2}\phi({\bf \ell})e^{\imath {\boldsymbol \theta}\cdot \hat{\bf n}} \,.
\end{equation}
The relation between the convergence and lensing kernel \eqref{eqn:kernel_kappa} is then
\begin{equation} \label{eqn:kernel_kappa2}
    I_\ell^{\kappa}(k) \simeq \frac{\ell^2}{2} I_\ell^{\phi}(k) \,.
\end{equation}
The flat-sky approximation does not affect the {\em ratio} since the difference in the prefactor 
$\ell+1/2 \to \ell$ cancels out.

\subsubsection{Narrow foreground}

If the redshift distribution of the foreground population is narrow enough in redshift or 
if we have a redshift accuracy $\sigma_z$ sufficient to slice the foreground population in 
narrow redshift bins, we can approximate the foreground redshift distribution as a Dirac delta-function
\begin{equation}
    W_f(\chi) \propto \delta_{\rm D}(\chi-\chi_f) \,,
\end{equation}
where $\chi_f$ is the peak of the distribution. We then find
\begin{multline} 
\label{eqn:phiG_narrow}
    C_\ell^{\phi G}(z_f,z_b) = \frac{2}{(\ell+1/2)^2} \left(\frac{3\Omega_{\rm m}H_0^2}{2c^2}\right) \\ \times
        \frac{b(\chi_f) P_\delta\left(\frac{\ell+1/2}{\chi_f},\chi_f\right)}{a(\chi_f)\chi_f}
        \int \dd \chi \frac{\chi-\chi_f}{\chi}W_b(\chi) \,.
\end{multline}
Under these approximations the ratio loses the $\ell$-dependence, we obtain a quantity which depends only 
on background parameters ($H_0$, $\Omega_X$, $w_0$, ...), and the clustering bias cancels out, i.e.
\begin{equation} \label{eqn:ratio_approx}
    r = \frac{\chi_*-\chi_f}{\chi_*} \frac{1}{\int \dd \chi \frac{\chi-\chi_f}{\chi}W_b({\chi})} \,.
\end{equation}

Finally, if also the background distribution is sufficiently thin we can recover the standard 
cosmographic expression for the ratio
\begin{equation} \label{eqn:ratio_approx}
    r = \frac{\chi_*-\chi_f}{\chi_b-\chi_f} \frac{\chi_b}{\chi_*} \,,
\end{equation}
where we assumed $W_b(\chi) \propto \delta_{\rm D}(\chi-\chi_b)$.

\subsection{Forecasts for future experiments}
We quantify the accuracy that will be reachable on the lensing ratio measurement for future experiments in the limit in which it can be considered as an $\ell$-independent quantity ($r_\ell \simeq r$). For this, we follow the formalism by DS in order to compute the error on $r$.

The log-likelihood is defined as
\begin{equation}
\label{loglik}
\chi^2(r) = \sum_\ell \frac{Z_\ell^2}{\sigma^2(Z_\ell)}
\end{equation}
where $Z_\ell = C_\ell^{\kappa_{\rm CMB} G} - r C_\ell^{\kappa_{\rm gal}G}$. For the variance of $Z_l$ at a fiducial value of the ratio $r_{0} $, we use the extended definition by \cite{Prat:2018yru}, which accounts for partial overlap in the sky between surveys,
\begin{multline}
\label{eq:sigmazl}
\sigma^2(Z_\ell) = \frac{1}{(2\ell+1)} \\ \times \bigg[ \frac{1}{f_{\rm sky}^{\kappa_{\rm CMB} G}} \Big(\bar{C}_\ell^{\kappa_{\rm CMB} \kappa_{\rm CMB}} \bar{C}_\ell^{GG} + (C_\ell^{\kappa_{\rm CMB} G})^2 \Big)\\
+ \frac{r_{0} ^2}{f_{\rm sky}^{\kappa_{\rm gal} G}} \Big( \bar{C}_\ell^{\kappa_{\rm gal} \kappa_{\rm gal}} \bar{C}_\ell^{GG} + (C_\ell^{\kappa_{\rm gal}G})^2 \Big) \\
- 2 r_{0} \frac{f_{\rm sky}^{\kappa_{\rm CMB} \kappa_{\rm gal} G}}{f_{\rm sky}^{\kappa_{\rm CMB}  G} f_{\rm sky}^{ \kappa_{\rm gal} G}} \Big( C_\ell^{\kappa_{\rm CMB} \kappa_{\rm gal}} \bar{C}_\ell^{GG} + C_\ell^{\kappa_{\rm CMB} G} C_\ell^{\kappa_{\rm gal} G} \Big) \bigg] 
\end{multline}
where $\bar{C}_\ell^{XX} = C_\ell^{XX} + {\cal N}_\ell^{XX}$ includes the signal and noise power spectra, and the $f_{\rm sky}$ factors account for the overlapping sky fraction between each pair of probes. We introduce the maximum likelihood estimator for the lensing ratio solving $\partial \chi^2/\partial r = 0$ as DS
\begin{equation}
\label{eq:rmax}
    \hat{r} = \frac{\sum_\ell C_\ell^{\kappa_{\rm CMB} G} C_\ell^{\kappa_{\rm gal} G}/\sigma^2(Z_\ell)}{\sum_\ell ( C_\ell^{\kappa_{\rm gal} G})^2/\sigma^2(Z_\ell)}\,,
\end{equation}

and we then compute the error on $\hat{r}$ as
\begin{equation}
\label{eq:sigmarell}
\frac{1}{\sigma^2(\hat{r})} = \frac{1}{2} \frac{\partial^2 \chi^2(r)}{\partial r^2}=  \sum_\ell \frac{( C_\ell^{\kappa_{\rm gal} G})^2}{\sigma^2 (Z_\ell)} \,.
\end{equation}

\begin{figure}
    \centering
    \includegraphics[width = \columnwidth]{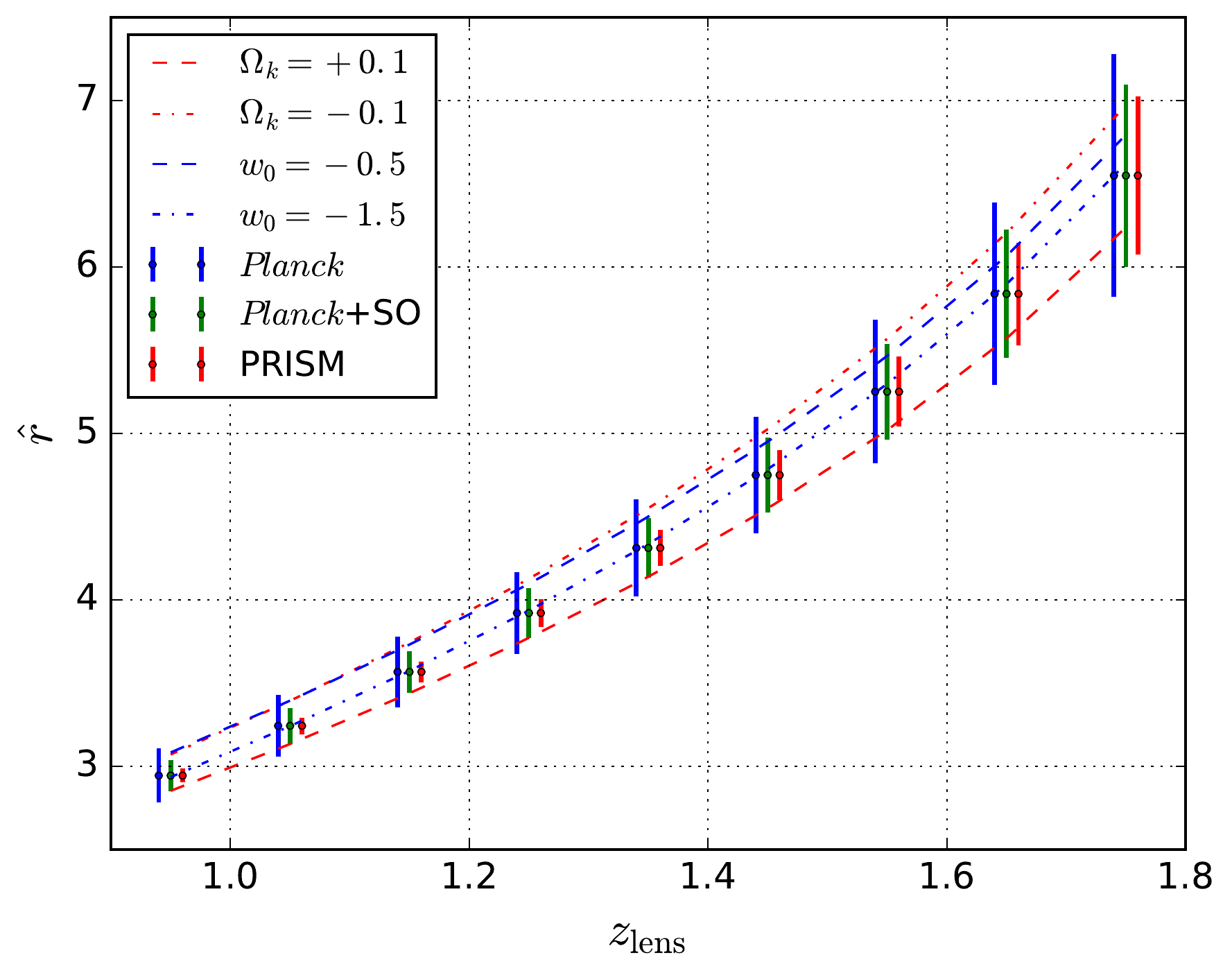}
    \caption{Forecast measurements of the maximum likelihood estimator for the lensing ratio $\hat{r}$ \eqref{eq:rmax} with their errors as a function of the  Euclid-like spectroscopic foreground redshift $z_{\rm lens}$ for three CMB experiments: $Planck$, $Planck$+SO and PRISM. The central dots and errorbars correspond to a $\Lambda$CDM cosmology while the red and blue dashed curves represent the values of $\hat{r}$ obtained shifting by a certain amount $\Omega_k$ and $w_0$, respectively.} 
    \label{fig:rhat}
\end{figure}

In Fig.~\ref{fig:rhat} we represent $\hat{r}$ and its error as a function of $z_{\rm lens}$ for the 9 possible bins of the Euclid-like spectroscopic survey and three of the CMB experiments considered: $Planck$, $Planck$+SO and PRISM. As suggested in \cite{Das:2008am,Prat:2018yru}, this estimator is specially sensitive to the curvature of the Universe and the equation of state of dark energy. We therefore calculate also $\hat{r}$ for cosmologies beyond $\Lambda$CDM shifting $w_0$ and $\Omega_k$ by a given amount. 

\begin{figure}
    \centering
    \includegraphics[width = \columnwidth]{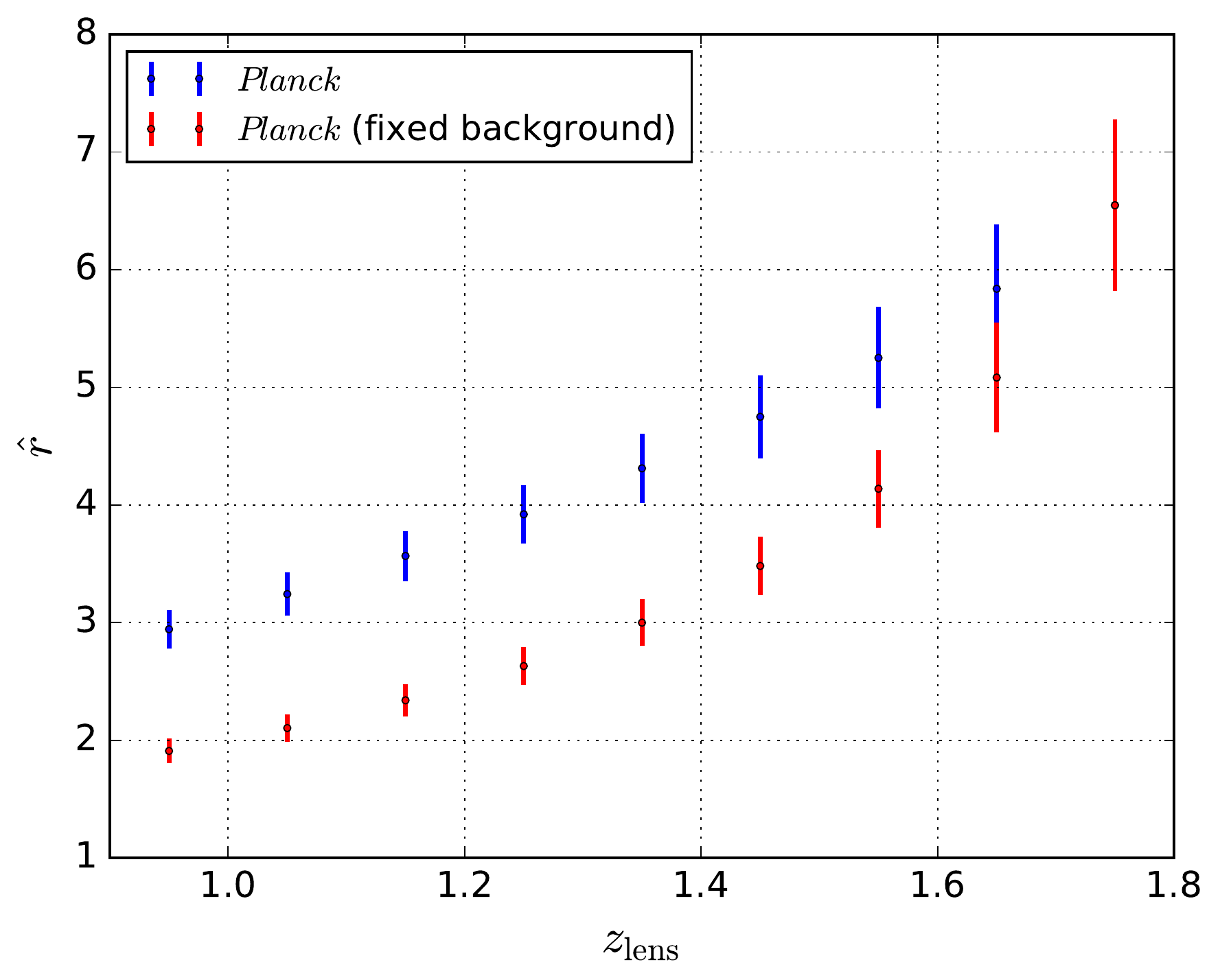}
    \caption{As for Fig.~\ref{fig:rhat} but using $Planck$ as CMB lensing background with a variable galaxy background beyond each foreground (blue errorbars) and a fixed galaxy background at 2.0 $< z <$ 2.5 (red errorbars).}
    \label{fig:rhatfixed}
\end{figure}

We find that for post-$Planck$ CMB experiments in which the CMB lensing noise will be reduced by a significant amount, the ratio will be measured with better accuracy specially for the lower redshift bins. At higher redshift for the lenses $z_{\rm lens}$, the higher noise of the galaxy surveys gives less precise measurements. For the non-standard cosmologies, we find that $\hat{r}$ is sensitive in particular to the curvature variations. 

For the Euclid-like spectroscopic lenses and using the $Planck$ CMB lensing, the best measurement will be $\sigma(\hat{r})/\hat{r}$ = 5.5\%, corresponding to the Euclid-r1 configuration at 0.9 $< z_{\rm lens} <$ 1.0. If we take a lower redshift lens at 0.2 $< z_{\rm lens} <$ 0.3 for DESI and SPHEREx, we get as relative errors 6.7\% and 4.3\%, respectively. This means that the measurement for DESI will be affected by the small overlapping sky fraction with Euclid, while using SPHEREx as foreground population can relatively improve the lensing ratio measurement. Using post-$Planck$ CMB lensing, we get as relative errors 3.2\% and 2.3\% for SO in combination with the Euclid-r1 and SPHEREx-r1 configurations, respectively. With PRISM, these two measurements improve to 1.4\% and 0.7\%, respectively.

In Fig.~\ref{fig:rhatfixed} we explore the effect of fixing the background galaxy shear sources to the Euclid-like photometric population placed behind the Euclid-like spectroscopic survey (i.e. behind the higher redshift lens on Fig.~\ref{fig:rhat}). The increase on the distances between the galaxy lens and source planes shifts the maximum likelihood ratio to lower values and also decreases the absolute error, but we find a very similar relative error in comparison with the variable background case. This also shows that the measurement is not limited by the background noise.

\section{A generalized lensing ratio estimator}
\label{sec:generalized}
In this section we show how the inclusion of the RSD and lensing magnification contributions to the galaxy number counts has an impact on the angular scale dependence of the lensing ratio $r_\ell$ and the cosmological information contained on it. We propose the introduction of a multipole dependent estimator to upgrade the formalism by DS and consider a more general case beyond assuming that $r$ is constant. We define the signal-to-noise ratio of the $\hat{r}_\ell$ estimator and evaluate the impact of including on the calculation the contributions beyond the density term.

\subsection{Number counts angular power spectrum}

Eq.~\eqref{eqn:kernel_counts} assumes only the contribution from the synchronous-gauge galaxy 
overdensity to the galaxy number counts. Here we quantify the relevance of including other terms 
to $\Delta^s_{\ell}(k,\chi)$ given by RSD and lensing magnification 
(see \cite{Challinor:2011bk,Ballardini:2018cho} for details). 
The RSD term is given by
\begin{equation}
    \Delta^{\rm RSD}_\ell (k,\chi) = \frac{k v_k}{\cal H} j_\ell''(k\chi) 
\end{equation}
where $v_k$ is the velocity of the sources and ${\cal H}$ is the Hubble parameter. The lensing convergence contribution is given by

\begin{multline}
\label{eq:deltalens}
\Delta^{\rm lensing}_\ell (k,\chi) =\frac{\ell(\ell+1)}{2} (2-5s) \\
 \times
\int_0^\chi {\rm d}\chi' \frac{\chi-\chi'}{\chi\chi'}
\left[\phi_k(\chi')+\psi_k(\chi')\right]j_\ell(k\chi')
\end{multline}

where $\phi_k$ and $\psi_k$ are the metric perturbations in the longitudinal gauge 
and $s$ is the magnification bias, which accounts for the fact that observed galaxies are magnified by lensing. In this paper, we consider lensing magnification as the only 
observational effect on number counts with the density and RSD. 
We neglect the Doppler, Sachs-Wolfe and other integrated effects (ISW and time-delay) because they 
are negligible in the calculation of the {\em ratio}.

 We derive and fit the functional form of the redshift dependence of the magnification bias $s(z)$ of the Euclid-like spectroscopic survey using the {\em model 3} luminosity function by \cite{Pozzetti:2016cch}. For a flux threshold of $F_{\rm cut} = 2 \times 10^{-16}$ erg s$^{-1}$ cm$^{-2}$ we find:
 
 \begin{equation}
     \label{eqn:magnification}
     s(z) = 0.33 + 0.46z+0.15z^2-0.16z^3+0.03z^4
 \end{equation}

Note that this fit is valid only for $z \geq 0.6$ since the {\em model 3} by  \cite{Pozzetti:2016cch} does not include data for low redshift objects.

\begin{figure*}
    \centering
    \includegraphics[width = \textwidth]{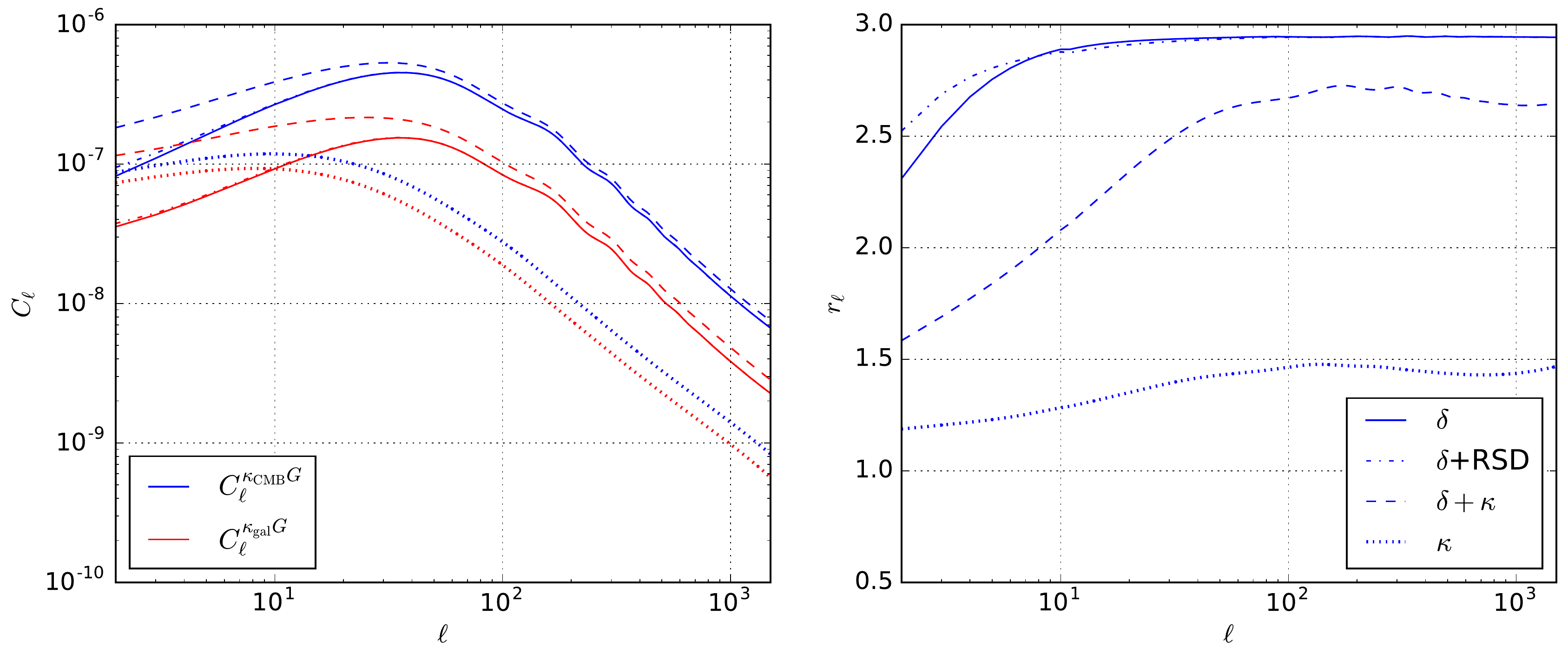}
    \caption{Impact of the RSD and lensing contributions on the lensing-galaxy cross-correlation angular power spectra (left panel) and on the lensing ratio estimator (right panel) for the Euclid-r1 configuration. }
    \label{fig:gr}
\end{figure*}

In Fig.~\ref{fig:gr} we show the impact of including the RSD term alone and both RSD and lensing contributions together in the cross-correlation angular power spectra $C_\ell^{\kappa_{\rm CMB} G}$ and $C_\ell^{\kappa_{\rm gal} G}$ and  in the lensing ratio $r_\ell$, adopting the Euclid-r1 configuration.

For the case including only RSD, we find a small correction on the angular power spectra at low multipoles, which results in a slightly higher lensing ratio at $\ell < 10$. When we consider also the lensing magnification, we obtain a larger positive contribution to both $C_\ell^{\kappa_{\rm CMB} G}$ and $C_\ell^{\kappa_{\rm gal} G}$ that is especially strong at large scales but holds also at higher multipoles. This results in a negative $\sim 10\%$ shift for $r_\ell$ given that the impact of including GR in the denominator $C_\ell^{\kappa_{\rm gal} G}$ is higher. The shape of the lensing ratio becomes less constant with $\ell$ once the lensing magnification contribution is considered.

We show here how the lensing term induces the $\ell$-dependence of the lensing ratio. If we consider the density term and the lensing contribution the kernel of the galaxy number counts becomes

\begin{multline} 
    I^G_\ell(k) \equiv I^{G_1}_\ell(k) + I^{G_2}_\ell(k) \\
    = \int \frac{\dd \chi}{(2\pi)^{3/2}} W_f(\chi)
    \left[\Delta^s_{\ell}(k,\chi) + \Delta^{\rm lensing}_\ell(k,\chi)\right]\,.
\end{multline}

Using the Limber approximation, the first term of the lensing-galaxy cross-correlation is given by Eq.~\eqref{eq:limberphiG}, and for the second term we find:

\begin{multline}
    C_\ell^{\phi G_2}(z_i,z_j) = \frac{\ell(\ell+1)}{(\ell+1/2)^2} 
        \left(\frac{3\Omega_{\rm m}H_0^2}{2c^2}\right) \\ \times
        \int \frac{\dd \chi}{(2\pi)^3} \frac{q_{b_i}(\chi)}{a(\chi)\chi} 
        \delta\left(\frac{\ell+1/2}{\chi},\chi\right)\left(\phi+\psi\right) \\ \times
        \int_0^\chi \dd \chi' \frac{\chi'-\chi}{\chi'\chi}(2-5s)W_{f_j}(\chi')\,.
\end{multline}
We note that in this case, the assumption of a narrow foreground would not eliminate the $\ell$-dependence of the ratio since the last integral is bound to $\chi$ and can not be simplified. We represent in Fig.~\ref{fig:gr} the contribution to the angular power spectra from the lensing magnification term ($\kappa$) and their ratio, showing that is not anymore an $\ell$-independent quantity. Nonetheless, the $\ell$-dependence can be alleviated using a different tracer for the galaxy foreground population which is not affected by the lensing magnification contribution as in \cite{Pourtsidou:2015qaa}, where they used as a foreground the SKA HI intensity mapping survey.

\subsection{Signal-to-noise analysis}
\label{sec:snr}
 We extend here the formalism to compute the error on the lensing ratio by DS to consider the angular scale dependence of the ratio. We introduce the signal-to-noise ratio (SNR) of an $\ell$-dependent estimator $\hat{r}_\ell$ and compare its value to the ratio studied before.

We start by assuming that different multipoles $\ell$ are uncorrelated. This is a consequence of neglecting the super-sample covariance and non-Gaussian terms of the covariance matrix of the data, therefore the remaining Gaussian term, which is diagonal in $\ell$, is assumed to be dominant at the scales of interest. We then define the log-likelihood of $r_\ell$ as
\begin{equation}
\label{loglik}
\chi_\ell^2(r_\ell) = \frac{Z_\ell^2}{\sigma_\ell^2(Z_\ell)}
\end{equation}
where $Z_\ell = C_\ell^{\kappa_{\rm CMB} G} - r_\ell C_\ell^{\kappa_{\rm gal}G}$. For the variance $\sigma_\ell(Z_\ell)$ we 
extend the definition in Eq.~\eqref{eq:sigmazl} replacing $r_0$ by a multipole dependent fiducial $r_{\ell,0}$

\begin{multline}
\label{sigmaZl}
\sigma_\ell^2(Z_\ell) = \frac{1}{(2\ell+1)} \\ \times \bigg[ \frac{1}{f_{\rm sky}^{\kappa_{\rm CMB} G}} \Big(\bar{C}_\ell^{\kappa_{\rm CMB} \kappa_{\rm CMB}} \bar{C}_\ell^{GG} + (C_\ell^{\kappa_{\rm CMB} G})^2 \Big)\\
+ \frac{r_{\ell,0} ^2}{f_{\rm sky}^{\kappa_{\rm gal} G}} \Big( \bar{C}_\ell^{\kappa_{\rm gal} \kappa_{\rm gal}} \bar{C}_\ell^{GG} + (C_\ell^{\kappa_{\rm gal}G})^2 \Big) \\
- 2 r_{\ell,0} \frac{f_{\rm sky}^{\kappa_{\rm CMB} \kappa_{\rm gal} G}}{f_{\rm sky}^{\kappa_{\rm CMB}  G} f_{\rm sky}^{ \kappa_{\rm gal} G}} \Big( C_\ell^{\kappa_{\rm CMB} \kappa_{\rm gal}} \bar{C}_\ell^{GG} + C_\ell^{\kappa_{\rm CMB} G} C_\ell^{\kappa_{\rm gal} G} \Big) \bigg] \,.
\end{multline}
The maximum likelihood estimator for the lensing ratio, $\hat{r}_\ell$, is obtained imposing $\partial \chi_\ell^2(r_\ell)/\partial r_\ell = 0$ as

\begin{equation}
\label{eq:rmaxell}
    \hat{r}_\ell = \frac{ C_\ell^{\kappa_{\rm CMB} G} C_\ell^{\kappa_{\rm gal} G}/\sigma^2(Z_\ell)}{ ( C_\ell^{\kappa_{\rm gal} G})^2/\sigma^2(Z_\ell)} = \frac{C_\ell^{\kappa_{\rm CMB} G}}{C_\ell^{\kappa_{\rm gal} G}}\,,
\end{equation}
which in this case coincides with the definition of the lensing ratio itself. We then estimate the error on $\hat{r}_\ell$ as
\begin{equation}
\label{eq:sigmarell}
\frac{1}{\sigma_\ell^2(\hat{r}_\ell)} = \frac{1}{2} \frac{\partial^2 \chi^2_\ell(r_\ell)}{\partial r_\ell^2}=  \frac{( C_\ell^{\kappa_{\rm gal} G})^2}{\sigma_\ell^2 (Z_\ell)}
\end{equation}
and with this, we define the SNR of the lensing ratio as the total one as sum over the multipoles since they are uncorrelated, hence 

\begin{equation}
\label{eq:snr}
\Big(\frac{S}{N}\Big)^2_{\hat{r}_\ell} = \sum_\ell
\frac{\hat{r}_\ell^2}{\sigma_\ell^2(\hat{r}_\ell)} = \sum_\ell \frac{( C_\ell^{\kappa_{\rm CMB} G})^2}{\sigma_\ell^2 (Z_\ell)} \,.
\end{equation} 

\begin{figure}
    \centering
    \includegraphics[width = \columnwidth]{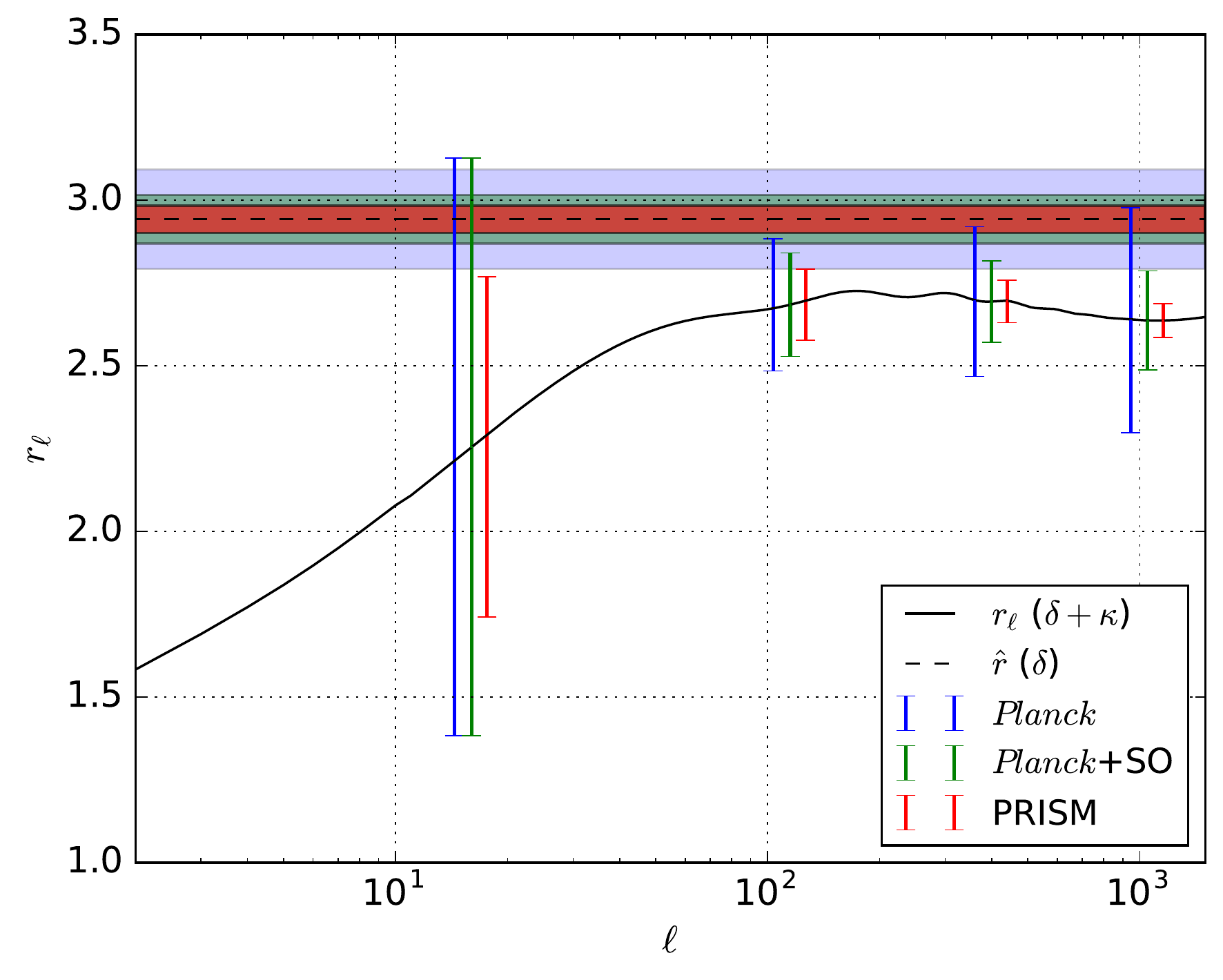}
    \caption{Lensing ratio and its error measured for the lowest $z_{\rm lens}$ configuration of the Euclid-like spectroscopic survey. The solid black curve represents $r_\ell$ computed with all the contributions to the galaxy number counts and the dashed line to $\hat{r}$ computed using the density term only. The blue, green and red error bars correspond errors on $r_\ell$ for $Planck$, $Planck$+SO and PRISM, respectively, in four broad bins - i.e. $(2,29), (30,199), (200,599), (600,1500)$. The blue, green and red shaded areas represent the 1$\sigma$ confidence region for $\hat{r}$ calculated for $Planck$, $Planck$+SO and PRISM, respectively.}
    \label{fig:detection}
\end{figure}

We now assess whether the $\ell$-dependence of the lensing ratio will be measurable using the future experiments discussed here. We find that this dependence will be detectable at the level of 1.4$\sigma$ for $Planck$, 2.2$\sigma$ for $Planck$+SO and 5.1$\sigma$ for PRISM by considering unbinned multipoles. We also visualize the importance of the multipole dependence in Fig.~\ref{fig:detection}, where we display 
for the Euclid-r1 configuration the errors on $r_\ell$ for $Planck$, $Planck$+SO and PRISM, by taking as an example 4 broad bins - see caption for more details. 
%in $\ell$, i.e. $(2,29), (30,199), (200,599), (600,1500)$.
Both the calculation with unbinned multipoles and Fig.~\ref{fig:detection} show how the multipole dependence induced by lensing magnification could be detected with future experiments.
%We take the Euclid-r1 configuration and calculate the errors on $r_\ell$ for $Planck$, $Planck$+SO and PRISM using 4 broad bins in $\ell$. The chosen bin edges are: ($\ell_{\rm min}$,$\ell_{\rm max}$) = (2,29), (30,199), (200,599), (600,1500). %\red{This choice is arbitrary and does not modify the final result about the $\ell$-dependence measurement, since the uncertainties from all the bins are added in a quadratic sum and we have assumed that the multipoles are uncorrelated.} 
%We represent the result in Fig.~\ref{fig:detection}, and we then estimate whether the angular scale dependence induced by the lensing contribution to the number counts will be detectable with future experiments, taking into account the errors on $\hat{r}$ and the errors for the $r_\ell$ bins. 
%Assuming that the multipole bins are uncorrelated, we get this dependence will be detectable at the level of 1.8$\sigma$ for $Planck$, 2.9$\sigma$ for $Planck$+SO and 6.3$\sigma$ for PRISM. 

\begin{figure*}
    \includegraphics[width=\textwidth]{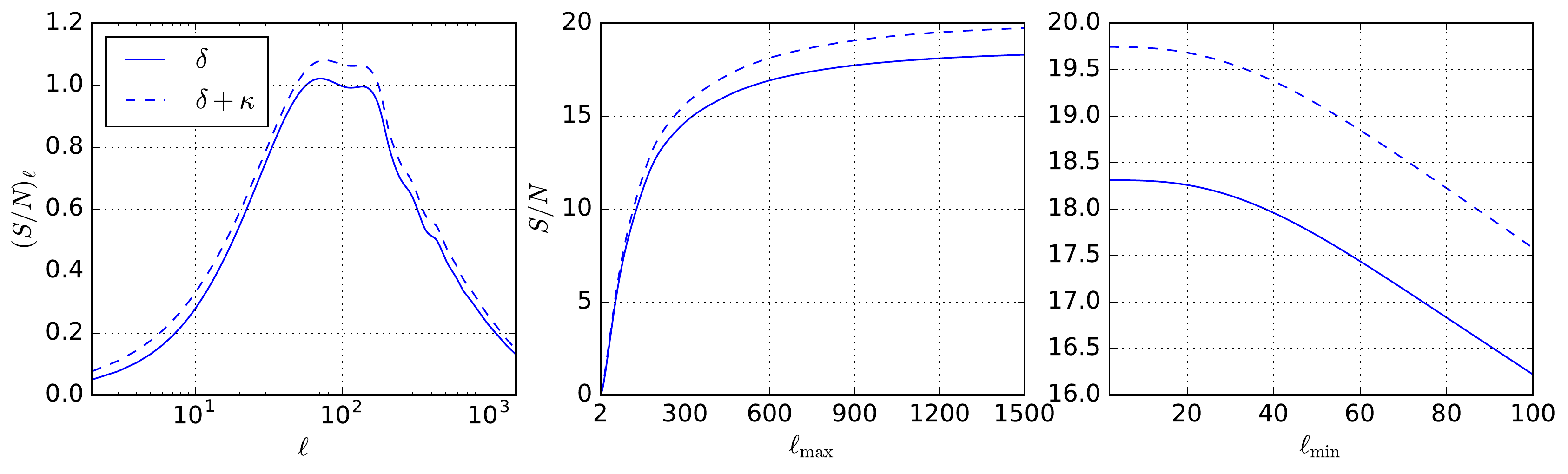}
    \caption{Contribution of each $\ell$ to the signal-to-noise of $r_\ell$ (left panel) and cumulative signal-to-noise as a function of $\ell_{\rm max}$ (center panel) and $\ell_{\rm min}$ (right panel). The solid lines represent the case including only the density term ($\delta$) and the dashed lines correspond to the calculation including also the lensing magnification term ($\delta + \kappa$). }
    \label{fig:snr_ell}
\end{figure*}

We calculate and show in Fig.~\ref{fig:snr_ell} the SNR of the lensing ratio for the Euclid-r1 configuration as a function of $\ell_{\rm max}$ and $\ell_{\rm min}$, as well as the individual contribution of each multipole to the total amount. We find that the majority of the information of this estimator is around $\ell \sim 100$. We also find that including corrections from general relativity increases the SNR around $\lesssim 10\%$.

In Tab.~\ref{tab:error} we list the SNR of the lensing ratio measurement for the CMB experiments considered as a function of the lens redshift $z_{\rm lens}$ of each bin. We find that the post-$Planck$ CMB lensing will reduce significantly the error on this measurement, reaching up to $\sim 1\%$ with PRISM. We also note the synergies between the Euclid-like survey and the CMB space missions due to the overlapping sky fraction, as an example PICO will be able to measure the lensing ratio with better accuracy than S4 despite having a larger CMB lensing noise. The impact of the lensing correction is stronger at higher $z_{\rm lens}$, allowing to increase the SNR up to a factor $\sim$ 2-3 for the last redshift bin. We have also checked that complementing the ground-based SO and S4 experiments with $Planck$ at $\ell < 30$ does not have a significant impact on the SNR.

%We now extend the SNR calculation to a tomographic case where we combine the various ratios, which can be useful in order to achieve better constraints for the cosmological parameters. Since for a Euclid-like experiment there is overlap between the backgrounds of the 9 possible ratios and also between some foregrounds and backgrounds, we take into account the covariance of each pair of ratios and define the combined SNR of $N$ lensing ratios as
%\begin{equation}
%\label{eq:snrcomb}
%\Big(\frac{S}{N}\Big)^2_{\rm comb} = \sum_{i,j}^{N} \sum_\ell
%\hat{r}_\ell^i [{\rm Cov}(\hat{r}_\ell)]^{-1}_{ij} \hat{r}_\ell^j \,,
%\end{equation} 
%where the $i$, $j$ indices stand for the ratios and% similarly to Eq.~\eqref{eq:sigmarell} we introduce the covariance of $Z_\ell$ as
%\begin{equation}
%\label{eq:covrl}    
%[{\rm Cov}(\hat{r}_\ell)]^{-1}_{ij} = C_\ell^{\kappa_{\rm gal}^i G^i} [{\rm Cov}(Z_\ell)]^{-1}_{ij} C_\ell^{\kappa_{\rm gal}^j G^j}\,,
%\end{equation}
 
%We list in Tab.~\ref{tab:error} ...

\begin{table}
%\centering
    \begin{tabular}{|l|cc|cc|cc|cc|cc|cc|}
    \hline 
    
    \multirow{2}{*}{$z_{\rm lens}$} & \multicolumn{2}{c|}{$Planck$}  &
    \multicolumn{2}{c|}{LiteBIRD}  &\multicolumn{2}{c|}{SO}  &
    \multicolumn{2}{c|}{S4}  &
    \multicolumn{2}{c|}{PICO} &
     \multicolumn{2}{c|}{PRISM}\\ 
    & $\delta$ & $\delta$+$\kappa$ 
   & $\delta$ & $\delta$+$\kappa$ &
   $\delta$ & $\delta$+$\kappa$ & $\delta$ & $\delta$+$\kappa$ & $\delta$ & $\delta$+$\kappa$  & $\delta$ & $\delta$+$\kappa$ \\ \hline
    0.95 & 18 & 20 & 23 & 25 & 31 & 34 & 55 & 61 & 60 & 67 & 72 & 81 \\
    
    1.05 & 18 & 20 & 23 & 25 & 30 & 34 & 51 & 59 & 55 & 64 & 64 & 77 \\    
    
    1.15 & 17 & 19 & 22 & 25 & 28 & 33 & 46 & 56 & 50 & 62 & 56 & 72 \\       
  
    1.25 & 16 & 19 & 20 & 25  & 26 & 33 & 40 & 53 & 43 & 58 & 48 & 67 \\   
    
    1.35 & 15 & 19 & 19 & 24 & 23 & 32 & 35 & 50 & 37 & 55 & 40 & 61 \\       
    
    1.45 & 14 & 19  & 17 & 24 & 21 & 30 &29 & 46 & 30 & 50 & 32 &  54 \\       
 
    1.55 & 12 & 18 & 15 & 23 & 18 & 28  & 23 & 42 & 24 & 45 & 25 & 48 \\   
    
    1.65 & 11 & 17 & 12 & 22 & 15 &  27 & 18 & 37 & 19 & 40 & 19 & 42 \\        
    
    1.75 & 9 & 16 & 10 & 20 & 12 &  25 & 13 & 32 & 14 & 35 & 14 & 36 \\       
   
    \hline
    \end{tabular}
    \caption{SNR of the lensing ratio for the CMB experiments considered as a function of the foreground redshift $z_{\rm lens}$ for the 9 bins of the Euclid-like spectroscopic survey. We list the SNR calculated using the density term only ($\delta$) and considering also the contribution from lensing magnification ($\delta$+$\kappa$).}
    \label{tab:error}
\end{table}

\section{Cosmological parameter constraints}
\label{sec:cosmoforecast}
We investigate here by a Fisher matrix approach whether the measurement of the lensing ratio can help to constrain cosmological parameters in extended models when it is added to the CMB information.

%\begin{equation}
%\label{eq:fisher1}
%    {\cal F}_{ij}^{\hat r} = \frac{\partial \hat{r}}{\partial p_i} \frac{1}{\sigma^2(\hat{r})} \frac{\partial \hat{r}}{\partial p_j} 
%\end{equation}
The Fisher matrix formalism \cite{astro-ph/9603021} assumes the likelihood $\cal{L}$ to be a multivariate Gaussian and the minimum errors on the cosmological parameters can be estimated from the diagonal of the inverse Fisher matrix ($\sigma_i \geq \sqrt{({\cal F}^{-1})_{ii}}$). We define the Fisher matrix of the lensing ratio $r_\ell$ as

\begin{equation}
\label{eq:fisher}
    {\cal F}_{\alpha \beta}^{r_\ell} \equiv \left\langle \frac{\partial^2 {\cal L}}{\partial \theta_\alpha \partial \theta_\beta} \right\rangle = \sum_\ell \frac{\partial r_\ell }{\partial \theta_\alpha} \frac{1}{\sigma_\ell^2(r_\ell)} \frac{\partial r_\ell }{\partial \theta_\beta} \,,
\end{equation}
where $\theta_\alpha$, $\theta_\beta$ are the cosmological parameters. The lensing ratio Fisher matrix is added as uncorrelated to the CMB Fisher matrix \cite{Jungman:1995bz,Eisenstein:1998hr}, which is given by
\begin{equation}
    \label{eq:cmbfisher} 
    {\cal F}_{\alpha \beta}^{\rm CMB}= \sum_\ell \frac{2\ell+1}{2} f_{\rm sky}^{\rm CMB} {\rm Tr} \left[  \frac{\partial {\cal C}}{\partial \theta_\alpha} {\cal C}^{-1} \frac{\partial {\cal C}}{\partial \theta_\beta}  {\cal C}^{-1} \right]\,,
\end{equation}

where $\cal{C}$ is the 3x3 covariance matrix of the CMB data including temperature ($TT$), polarization ($EE$), lensing ($\phi \phi$) and their cross-correlations.

For the cosmological model, we extend the baseline $\Lambda$CDM+$\sum m_\nu$ cosmology to a 9 parameter model where we allow also to vary the dark energy equation of state and the curvature density ($w_0$CDM+$\sum m_\nu$+$\Omega_k$), since we have shown in Section~\ref{sec:DS} that the lensing ratio is sensitive to the variation of these parameters. We adopt as fiducial values $w_0$ = -1 and $\Omega_k$ = 0.

\begin{figure*}
    \centering
    \includegraphics[width = \textwidth]{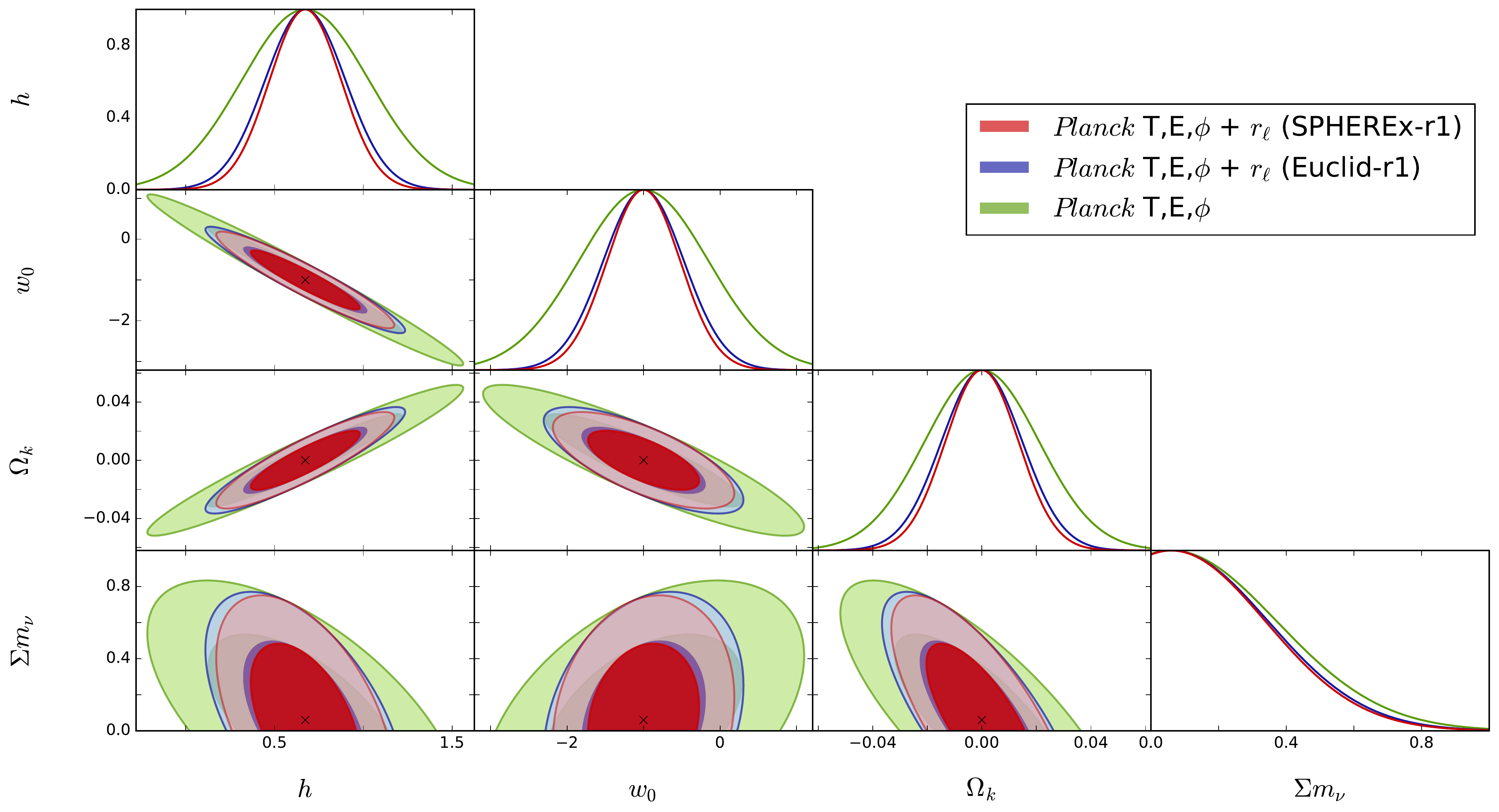}
    \caption{Marginalized 68\% and 95\% 2D confidence regions for a $w_0$CDM+$\sum m_\nu$+$\Omega_k$ model obtained by the Fisher matrix of a $Planck$-like experiment including temperature, polarization and lensing (green contours), and adding the Fisher matrix of the lensing ratio obtained using as lens population the first bin of the Euclid-r1 configuration at 0.9 $< z_{\rm lens} <$ 1.0 (blue contours) and the SPHEREx-r1 at 0.2 $ < z_{\rm lens} <$ 0.3 (red contours). We do not show the other 5 cosmological parameters since they are not sensitive to the addition of lensing ratio to the CMB information.}
    \label{fig:triangle_planck}
\end{figure*}

We show in Fig.~\ref{fig:triangle_planck} the 68\% and 95\% marginalized confidence regions for $h$, $w_0$, $\Omega_k$ and $\sum m_\nu$ obtained for a $Planck$-like CMB experiment following the Fisher matrix described in Eq.~\eqref{eq:cmbfisher} and the sum of both $r_\ell$ and CMB Fisher matrices. We calculate the $r_\ell$ Fisher matrix for the Euclid-r1 and SPHEREx-r1 configurations described in Section~\ref{sec:data}. The improvement found by adding the lensing ratio to the CMB information is about $\lesssim 40\%$ for $h$, $w_0$ and $\Omega_k$, while the neutrino mass is only marginally improved. For the spatial curvature we get a combined uncertainty of $\sigma(\Omega_k) \sim 0.015$, comparable to the $Planck$ 2018 \cite{1807.06209} error for a simpler $\Lambda$CDM+$\Omega_k$ model using CMB temperature and polarization. The constraints from the combination with the lensing ratio obtained with SPHEREx as foreground population are slightly better with respect to the Euclid-like spectroscopic lens. 

\begin{figure*}
    \includegraphics[width = \textwidth]{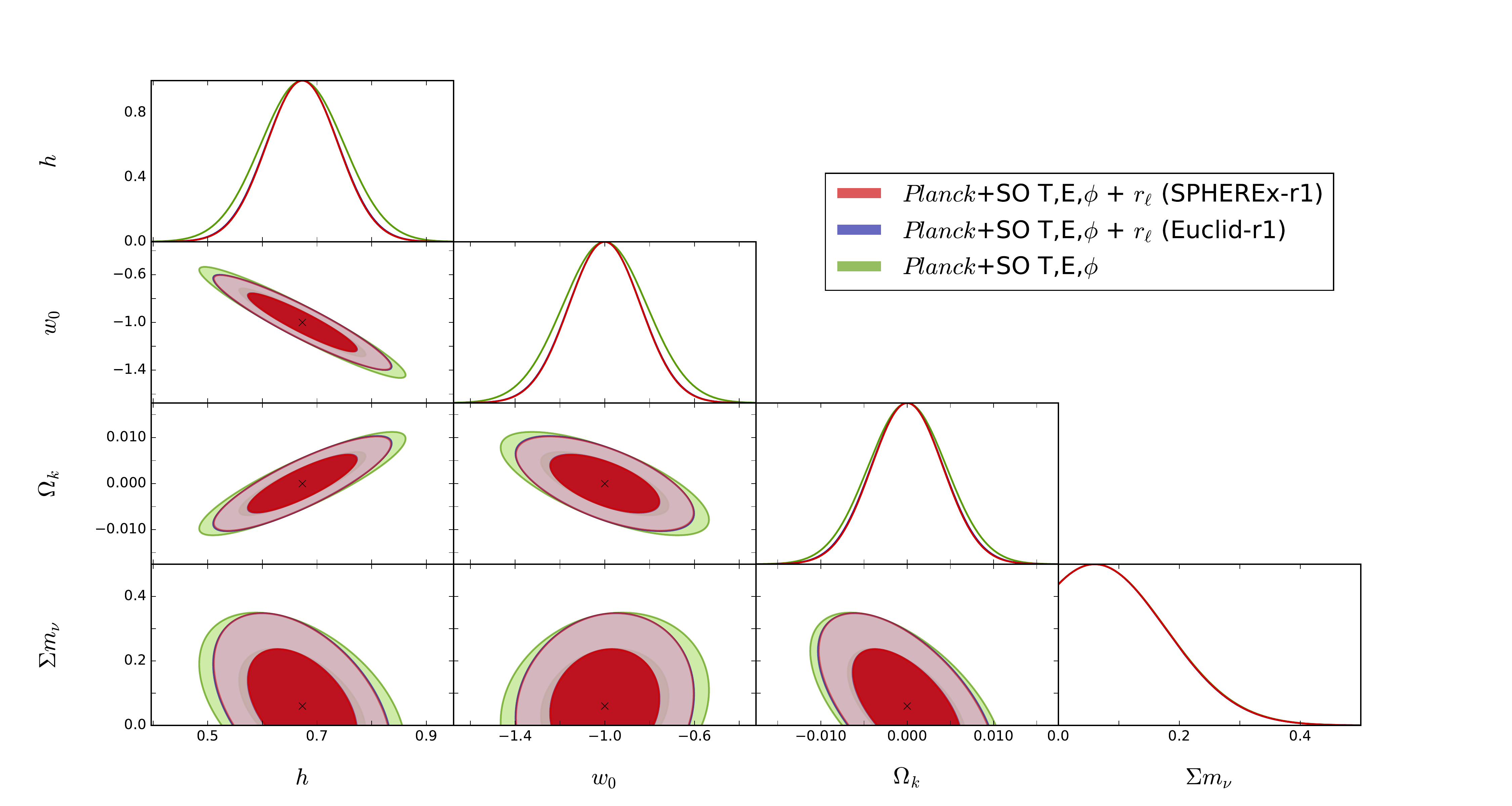}
    \caption{As for Fig.~\ref{fig:triangle_planck} but using $Planck$+SO as CMB experiment.}
    \label{fig:triangle_SO}
\end{figure*}

In Fig.~\ref{fig:triangle_SO} we show the same constraints but using $Planck$+SO as CMB experiment. For this case, we find relative improvements around $\sim 15\%$ for $h$ and $w_0$ and $\sim$ 10\% for $\Omega_k$ with respect to the CMB. For the spatial curvature error, we get $\sigma(\Omega_k) \sim
0.004$ for the combination of SO with Euclid-r1 or SPHEREx-r1.

\begin{figure*}
    \centering
    \includegraphics[width = \textwidth]{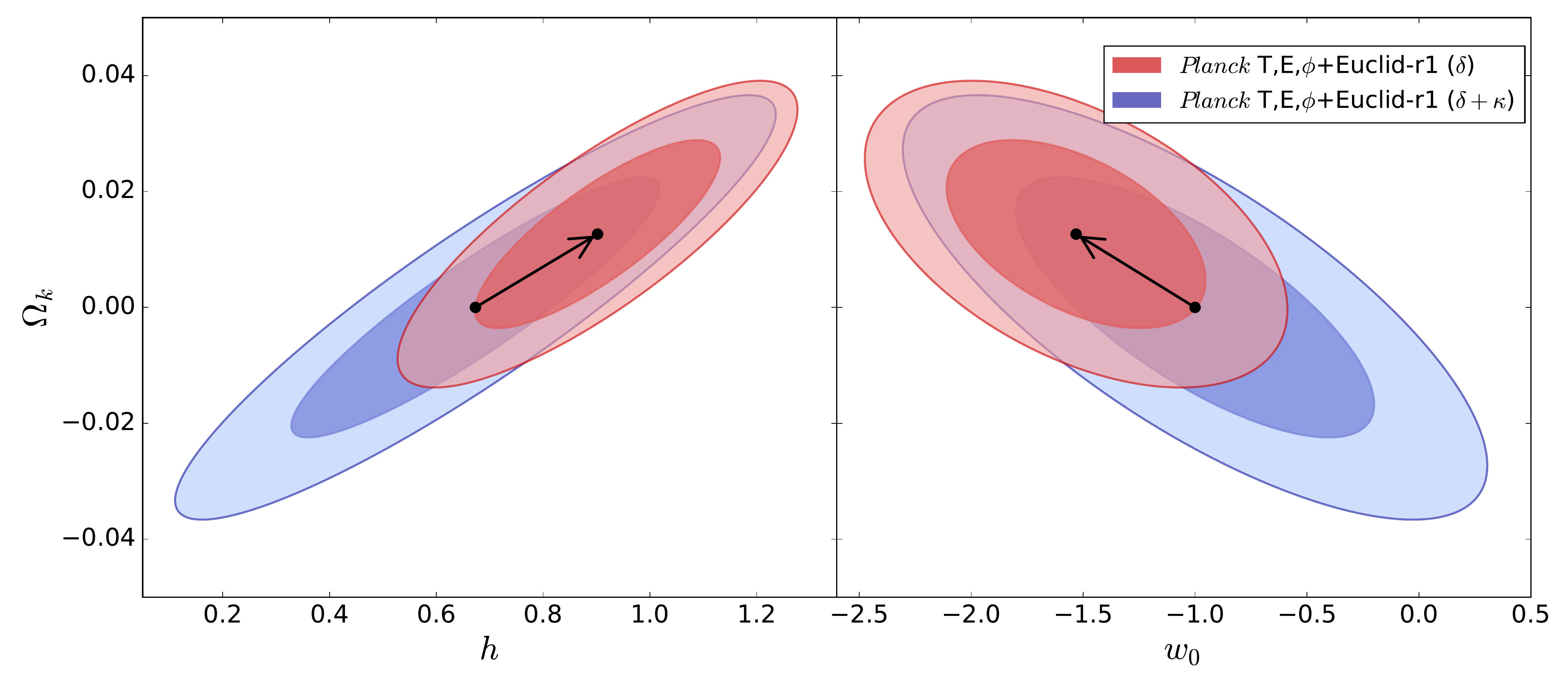}
    \caption{Predicted bias in the cosmological parameters induced by neglecting the lensing magnification contribution to the galaxy number counts in a combined analysis of the CMB and the lensing ratio, using $Planck$ and the Euclid-r1 configuration. The blue contours represent the marginalized 68\% and 95\% 2D confidence regions for the $h$-$\Omega_k$ and $w_0$-$\Omega_k$ planes obtained considering the lensing contribution ($\delta+\kappa$), while the red contours correspond to the case with the density term only ($\delta$).}
    \label{fig:bias}
\end{figure*}

We have shown that the $\hat{r}$ and $\hat{r}_\ell$ estimators -which neglect and include the contribution from lensing magnification, respectively- are different in terms of SNR. We now explore whether neglecting the inclusion of the lensing magnification term can induce a bias in the derived cosmological parameters.
Following the formalism by \cite{Kitching:2008eq}, it can be shown the predicted bias in the cosmological parameters due to an uncorrected effect/systematic is expressed as 
\begin{equation}
    b_{\theta_\alpha} =  (\widetilde{\cal F}^{-1})_{\alpha \beta} B_\beta \,,
\label{eq:bias}    
\end{equation}
where $\widetilde{\cal F}_{\alpha \beta}$ is the Fisher matrix computed by assuming the theoretical signal without the lensing magnification term. The vector $B_\beta$ is given by
\begin{equation}
    B_\beta = \sum_\ell \frac{1}{\sigma_\ell^{2}(\widetilde{r}_\ell)}(\widetilde{r}_\ell-r_\ell) \frac{\partial \widetilde{r}_\ell}{\partial \theta_\beta}\,,
\end{equation}
where $\widetilde{r_\ell}$ and $r_\ell$ are the lensing ratios obtained without and with the lensing magnification contribution, respectively. 

By our working assumptions, we compute the bias in the cosmological parameters for the combined constraints from $Planck$ and the lensing ratio using the Euclid-r1 configuration as an example. We represent the result in Fig.~\ref{fig:bias}, where we show the marginalized 68\% and 95\% 2D confidence regions for the $h$-$\Omega_k$ and $w_0$-$\Omega_k$ planes obtained considering and neglecting the lensing term, and we have shifted the uncorrected contours by the amount given by Eq.~\eqref{eq:bias}. We get for the bias on the parameters $b_{h} = 0.23$, $b_{w_0} = -0.53$ and $b_{\Omega_k} = 0.013$, which taking into account the uncertainties from both approaches corresponds to a shift of $0.85\sigma$ for $h$, $0.8\sigma$ for $w_0$ and $0.7\sigma$ for  $\Omega_k$. Whereas the forecast uncertainties by lensing ratio {\em alone} improve by adding lensing magnification, consistently with the improvement in the SNR shown in Tab.~\ref{tab:error}, it is clear from Fig.~\ref{fig:bias} that 
the forecast uncertainties in $h$-$\Omega_k$ and $w_0$-$\Omega_k$ in combination with the CMB degrade when taking into account lensing magnification. We interpret this effect as a consequence of introducing an improved lensing ratio which goes beyond its only dependence on distances and on background cosmology and therefore worsten the uncertainties on parameters as $h$-$w_0$-$\Omega_k$. Therefore, the neglection of lensing magnification term could  overestimate the constraints achievable with lensing ratios using lenses at the typical redshift of a Euclid-like spectroscopic survey, in which this contribution is important, and would lead to a potential bias in the cosmological parameters.

\begin{figure*}
    \centering
    \includegraphics[width = \textwidth]{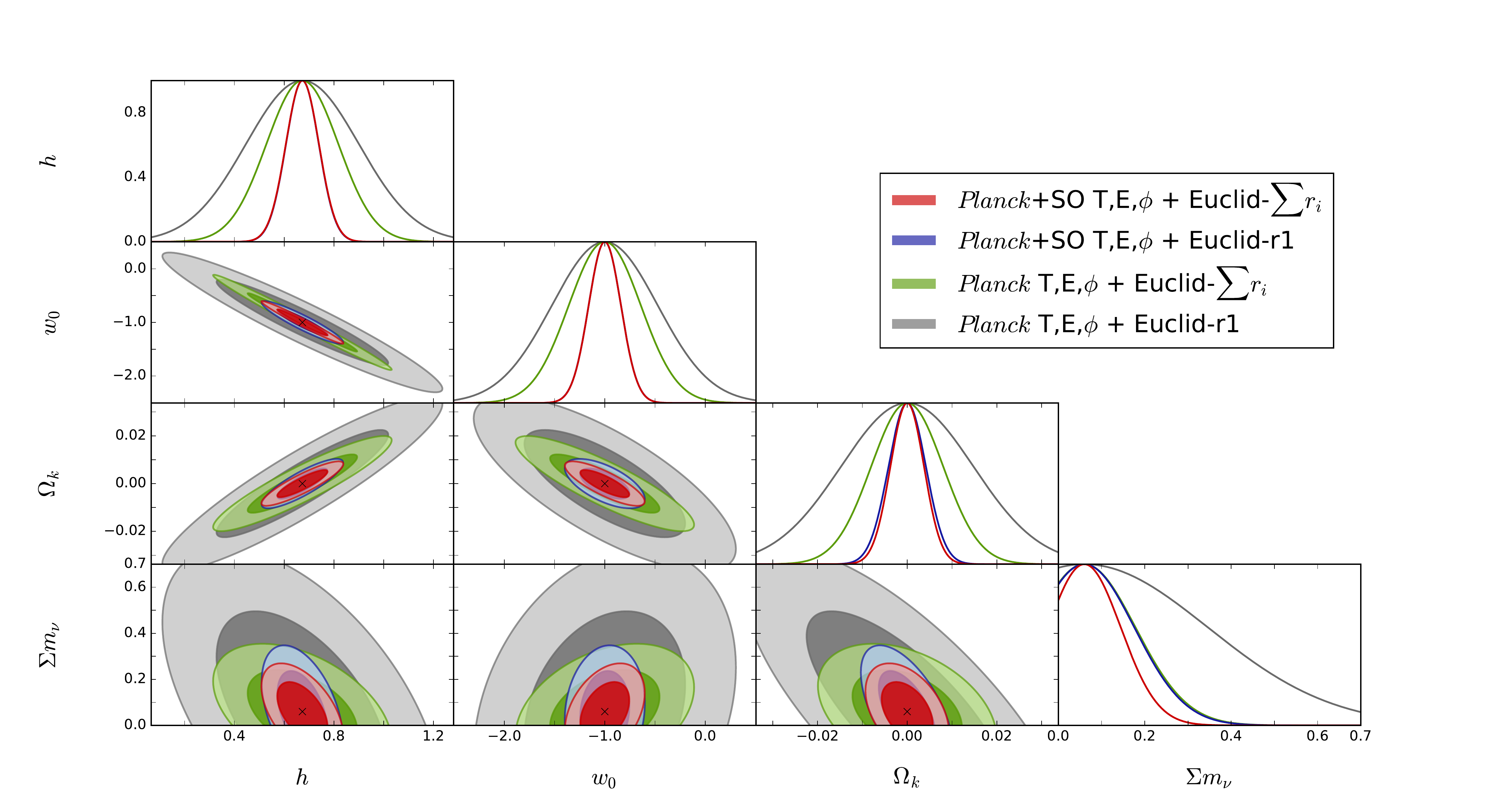}
    \caption{Marginalized 68\% and 95\% 2D confidence regions for a $w_0$CDM+$\sum m_\nu$+$\Omega_k$ model forecast by the combination of a $Planck$-like experiment with the Euclid-r1 lensing ratio configuration (grey contours) and the Euclid-$\sum r_i$ tomographic analysis (green contours), and by the combination of $Planck$+SO with the Euclid-r1 configuration (blue contours) and the Euclid-$\sum r_i$ tomographic analysis (red contours).}
    \label{fig:triangle_tomography}
\end{figure*}

We now explore the possibility of improving the cosmological parameter constraints by combining the 9 possible lensing ratio configurations for Euclid ranging from $z_{\rm lens} = 0.9$ to $z_{\rm lens} = 1.8$ in a joint tomographic measurement, which hereafter we call Euclid-$\sum r_i$. We introduce the covariance of $N$ ratios to take into account that in this combination there is redshift overlap between the different backgrounds and between some backgrounds and foregrounds. We then rewrite Eq.~\eqref{eq:fisher} as
\begin{equation}
\label{eq:fishertomography}
    {\cal F}_{\alpha \beta}^{r_\ell}  = \sum_\ell \sum_{i,j}^{N} \frac{\partial r_\ell^i }{\partial \theta_\alpha} 
   [ {\rm Cov}(\hat{r}_\ell)]^{-1}_{ij}
    \frac{\partial r_\ell^j }{\partial \theta_\beta} \,,
\end{equation}
where the $i,j$ indices run over the ratios and the elements of the covariance matrix Cov$(\hat{r}_\ell)_{ij}$ are given by
\begin{multline}
\label{covZl}
{\rm Cov}(\hat{r}_\ell)_{ij} = \frac{1}{(2\ell+1)} \frac{1}{ C_\ell^{\kappa_{\rm gal}^i G^i} C_\ell^{\kappa_{\rm gal}^j G^j}}\\ \times \bigg[ \frac{1}{f_{\rm sky}^{\kappa_{\rm CMB} G}} \Big(\bar{C}_\ell^{\kappa_{\rm CMB} \kappa_{\rm CMB}} \bar{C}_\ell^{G^iG^j} + C_\ell^{\kappa_{\rm CMB} G^i} C_\ell^{\kappa_{\rm CMB} G^j} \Big)\\
+ \frac{r_{\ell,0}^i r_{\ell,0}^j }{f_{\rm sky}^{\kappa_{\rm gal} G}} \Big( \bar{C}_\ell^{\kappa_{\rm gal}^i \kappa_{\rm gal}^j} \bar{C}_\ell^{G^i G^j} + C_\ell^{\kappa_{\rm gal}^i G^j} C_\ell^{\kappa_{\rm gal}^j G^i} \Big) \\
- r_{\ell,0}^i \frac{f_{\rm sky}^{\kappa_{\rm CMB} \kappa_{\rm gal} G}}{f_{\rm sky}^{\kappa_{\rm CMB}  G} f_{\rm sky}^{ \kappa_{\rm gal} G}} \Big( C_\ell^{\kappa_{\rm CMB} \kappa_{\rm gal}^i} \bar{C}_\ell^{G^i G^j} + C_\ell^{\kappa_{\rm CMB} G^i} C_\ell^{\kappa_{\rm gal}^i G^j} \Big)\\
- r_{\ell,0}^j \frac{f_{\rm sky}^{\kappa_{\rm CMB} \kappa_{\rm gal} G}}{f_{\rm sky}^{\kappa_{\rm CMB}  G} f_{\rm sky}^{ \kappa_{\rm gal} G}} \Big( C_\ell^{\kappa_{\rm CMB} \kappa_{\rm gal}^j} \bar{C}_\ell^{G^i G^j}\\ + C_\ell^{\kappa_{\rm CMB} G^j} C_\ell^{\kappa_{\rm gal}^j G^i} \Big) \bigg] \,.
\end{multline}

In Fig.~\ref{fig:triangle_tomography} we compare the constraints obtained for $Planck$ and $Planck$+SO in combination with the Euclid-r1 configuration to their combination with the tomographic Euclid-$\sum r_i$ measurement. For $Planck$, the constraints on $h$, $w_0$ and $\Omega_k$ are improved around $\sim 40 \%$ from the tomography with respect to the single ratio case, this corresponds to a $\sim 60$-70\% improvement with respect to $Planck$ alone. We get a joint uncertainty on the spatial curvature of $\sigma(\Omega_k) \sim 0.008$. The neutrino mass is the most benefited parameter from the tomography, reaching up to a $\sim 60 \%$ improvement with respect to the single bin case. For $Planck$+SO, the CMB has a higher relative weight but still the error on $\Omega_k$ is improved around $\sim 10\%$ with tomography, while the neutrino mass error is improved around $\sim$ 30\%.
\section{Conclusions}
\label{sec:conclusions}
We have studied the ratio between the galaxy number counts/CMB lensing and the galaxy number counts/galaxy shear cross-correlations as a cosmographic quantity \cite{Das:2008am}. We have forecast the scientific capabilities of this estimator using a Euclid-like experiment both for the galaxy background as will be measured from the photometric survey and for the galaxy foreground whose redshift will be determined from its spectroscopic survey, and $Planck$, LiteBIRD, SO, S4, PICO and PRISM for the CMB lensing background. A Euclid-like experiment could deliver tomographic measurements of the lensing ratio on the basis of the amount of lenses obtainable from the spectroscopic survey. We have then increased the lever arm in redshift by complementing the Euclid-like specifications with DESI and SPHEREx as galaxy foreground populations at lower redshift than $z_{\rm lens} = 0.9$
We have found that using SPHEREx as lens population and post-$Planck$ space missions as PRISM the lensing ratio will be measurable with a $\sim 0.7\%$ uncertainty.

We have also found a non-trivial angular scale dependence in the lensing ratio when going beyond the cosmographic limit \cite{Das:2008am}, i.e. when exact expressions are considered and relativistic corrections are taken into account. In particular, we show that the contribution from lensing magnification will be important for future experiments, as shown for other ratio estimators (see e.g. \cite{Dizgah:2016bgm,Ghosh:2018ijm}). Nonetheless, we show that the RSD contribution and the Limber approximation do not induce any significant effect.
We have found that this angular scale dependence of the lensing ratio 
will be especially important at higher redshift of the lenses and can be detectable at a statistical significant level with post-$Planck$ CMB lensing in combination with a Euclid-like experiment. The significance level could be increased in a tomographic analysis combining the lensing ratio measurements form different bins. This multipole dependence calls for the introduction of an ensemble of lensing ratios defined $\ell$ by $\ell$, with their corresponding optimal and minimum variance estimators, which we identify.
Using this new formalism, we have calculated the total signal-to-noise of the lensing ratio for a Euclid-like spectroscopic foreground including the contribution from the lensing magnification and compared it to the cosmographic limit approach. We have found an improvement in the signal-to-noise that ranges from $\sim 10\%$ to a factor $\sim$ 2-3, depending on the lens redsfhit. The majority of the information on this estimator is found to be around $\ell \sim 100$.

By using this improved estimator we forecast its capability
to constrain a non-flat cosmology with non-zero neutrino mass and a redshift-independent parameter of state for
dark energy in combination with future CMB experiments. We find that the inclusion of the lensing ratio can reduce by $\lesssim 40\%$ the uncertainties on $H_0$, $w_0$ and $\Omega_k$ from $Planck$. We also predict a non-negligible bias in the estimation of these cosmological parameters caused by neglecting the lensing magnification term in a combined analysis. We find that a Euclid-like experiment in combination with $Planck$ could provide a constraint on the spatial curvature with an uncertainty of $\sigma(\Omega_k) \sim 0.015$ for the first bin of the spectroscopic survey
centered at $z_{\rm lens}=0.95$. By considering a joint tomographic analysis of 9 lensing ratio measurements for a Euclid-like survey between $z_{\rm lens}=0.9$ and $1.8$, the uncertainty on the spatial curvature can be reduced to $\sigma(\Omega_k) \sim 0.008$ and we get a $\sim 60$-70\% improvement in the errors on $H_0$, $w_0$, $\Omega_k$ and $\sum m_\nu$ with respect to $Planck$. \\

\begin{acknowledgments}
We wish to thank Dhiraj Kumar Hazra for having sparked our interest in this topic and for collaboration at the initial stage of this work. We thank Jose Alberto Rubi\~no-Martin, Daniela Paoletti, Pauline Vielzeuf and David Spergel for useful discussions. 
We acknowledge partial financial contribution from the agreement ASI n.I/023/12/0 ``Attivit\`a relative
alla fase B2/C per la missione Euclid". We also acknowledge the support from the Ministero degli Affari Esteri della Cooperazione Internazionale - Direzione Generale per la Promozione del Sistema Paese Progetto di Grande Rilevanza 
ZA18GR02. MB was supported by the South African Radio Astronomy Observatory, which is a facility of the National Research Foundation, an agency of the Department of Science and Technology and by a Claude Leon Foundation fellowship. 
\end{acknowledgments}

% \clearpage
%%%%%%%%%%%%%%%%%%%%%%%%%%%%%%%%%%%%%%%%%%%%%%%%%%%%%%%%%%%%%%%%%%%%%%%%%%%%%%%

%%%%%%%%%%%%%%%%%%%%%%%%%%%%%%%%%%%%%%%%%%%%%%%%%%%%%%%%%%%%%%%%%%%%%%%%%%%%%%%

\begin{thebibliography}{99}
%\cite{Jain:2003tba}
\bibitem{Jain:2003tba} 
  B.~Jain and A.~Taylor,
  %``Cross-correlation tomography: measuring dark energy evolution with weak lensing,''
  Phys.\ Rev.\ Lett.\  {\bf 91}, 141302 (2003)
  doi:10.1103/PhysRevLett.91.141302
  [astro-ph/0306046].
  %%CITATION = doi:10.1103/PhysRevLett.91.141302;%%
  %350 citations counted in INSPIRE as of 19 Nov 2019


%\cite{Bernstein:2003es}
\bibitem{Bernstein:2003es} 
  G.~M.~Bernstein and B.~Jain,
  %``Dark energy constraints from weak lensing cross - correlation cosmography,''
  Astrophys.\ J.\  {\bf 600}, 17 (2004)
  doi:10.1086/379768
  [astro-ph/0309332].
  %%CITATION = doi:10.1086/379768;%%
  %116 citations counted in INSPIRE as of 19 Nov 2019

%\cite{Zhang:2003ii}
\bibitem{Zhang:2003ii} 
  J.~Zhang, L.~Hui and A.~Stebbins,
  %``Isolating geometry in weak lensing measurements,''
  Astrophys.\ J.\  {\bf 635}, 806 (2005)
  doi:10.1086/497676
  [astro-ph/0312348].
  %%CITATION = doi:10.1086/497676;%%
  %58 citations counted in INSPIRE as of 16 Dec 2019
  
%\cite{Bernstein:2005en}
\bibitem{Bernstein:2005en} 
  G.~Bernstein,
  %``Metric tests for curvature from weak lensing and baryon acoustic oscillations,''
  Astrophys.\ J.\  {\bf 637}, 598 (2006)
  doi:10.1086/498079
  [astro-ph/0503276].
  %%CITATION = doi:10.1086/498079;%%
  %57 citations counted in INSPIRE as of 19 Nov 2019


%\cite{Hu:2007jh}
\bibitem{Hu:2007jh} 
  W.~Hu, D.~E.~Holz and C.~Vale,
  %``CMB Cluster Lensing: Cosmography with the Longest Lever Arm,''
  Phys.\ Rev.\ D {\bf 76}, 127301 (2007)
  doi:10.1103/PhysRevD.76.127301
  [arXiv:0708.4391 [astro-ph]].
  %%CITATION = doi:10.1103/PhysRevD.76.127301;%%
  %18 citations counted in INSPIRE as of 19 Nov 2019


%\cite{Das:2008am}
\bibitem{Das:2008am} 
  S.~Das and D.~N.~Spergel,
  %``Measuring Distance Ratios with CMB-Galaxy Lensing Cross-correlations,''
  Phys.\ Rev.\ D {\bf 79}, 043509 (2009)
  doi:10.1103/PhysRevD.79.043509
  [arXiv:0810.3931 [astro-ph]].
  %%CITATION = doi:10.1103/PhysRevD.79.043509;%%
  %22 citations counted in INSPIRE as of 19 Nov 2019


%\cite{Ade:2018sbj}
\bibitem{Ade:2018sbj} 
  P.~Ade {\it et al.} [Simons Observatory Collaboration],
  %``The Simons Observatory: Science goals and forecasts,''
  JCAP {\bf 1902}, 056 (2019)
  doi:10.1088/1475-7516/2019/02/056
  [arXiv:1808.07445 [astro-ph.CO]].
  %%CITATION = doi:10.1088/1475-7516/2019/02/056;%%
  %152 citations counted in INSPIRE as of 19 Nov 2019


%\cite{Prat:2018yru}
\bibitem{Prat:2018yru} 
  J.~Prat {\it et al.} [DES and SPT Collaborations],
  %``Cosmological lensing ratios with DES Y1, SPT and Planck,''
  Mon.\ Not.\ Roy.\ Astron.\ Soc.\  {\bf 487}, no. 1, 1363 (2019)
  doi:10.1093/mnras/stz1309
  [arXiv:1810.02212 [astro-ph.CO]].
  %%CITATION = doi:10.1093/mnras/stz1309;%%
  %1 citations counted in INSPIRE as of 19 Nov 2019


%\cite{Miyatake:2016gdc}
\bibitem{Miyatake:2016gdc} 
  H.~Miyatake, M.~S.~Madhavacheril, N.~Sehgal, A.~Slosar, D.~N.~Spergel, B.~Sherwin and A.~van Engelen,
  %``Measurement of a Cosmographic Distance Ratio with Galaxy and Cosmic Microwave Background Lensing,''
  Phys.\ Rev.\ Lett.\  {\bf 118}, no. 16, 161301 (2017)
  doi:10.1103/PhysRevLett.118.161301
  [arXiv:1605.05337 [astro-ph.CO]].
  %%CITATION = doi:10.1103/PhysRevLett.118.161301;%%
  %15 citations counted in INSPIRE as of 19 Nov 2019


%\cite{1010.4915}
\bibitem{1010.4915} 
  M.~White {\it et al.},
  %``The clustering of massive galaxies at z~0.5 from the first semester of BOSS data,''
  Astrophys.\ J.\  {\bf 728}, 126 (2011)
  doi:10.1088/0004-637X/728/2/126
  [arXiv:1010.4915 [astro-ph.CO]].
  %%CITATION = doi:10.1088/0004-637X/728/2/126;%%
  %197 citations counted in INSPIRE as of 19 Nov 2019


%\cite{1210.8156}
\bibitem{1210.8156} 
  T.~Erben {\it et al.},
  %``CFHTLenS: The Canada-France-Hawaii Telescope Lensing Survey - Imaging Data and Catalogue Products,''
  Mon.\ Not.\ Roy.\ Astron.\ Soc.\  {\bf 433}, 2545 (2013)
  doi:10.1093/mnras/stt928
  [arXiv:1210.8156 [astro-ph.CO]].
  %%CITATION = doi:10.1093/mnras/stt928;%%
  %243 citations counted in INSPIRE as of 19 Nov 2019


%\cite{1502.01591}
\bibitem{1502.01591} 
  P.~A.~R.~Ade {\it et al.} [Planck Collaboration],
  %``Planck 2015 results. XV. Gravitational lensing,''
  Astron.\ Astrophys.\  {\bf 594}, A15 (2016)
  doi:10.1051/0004-6361/201525941
  [arXiv:1502.01591 [astro-ph.CO]].
  %%CITATION = doi:10.1051/0004-6361/201525941;%%
  %430 citations counted in INSPIRE as of 19 Nov 2019


%\cite{1708.01530}
\bibitem{1708.01530} 
  T.~M.~C.~Abbott {\it et al.} [DES Collaboration],
  %``Dark Energy Survey year 1 results: Cosmological constraints from galaxy clustering and weak lensing,''
  Phys.\ Rev.\ D {\bf 98}, no. 4, 043526 (2018)
  doi:10.1103/PhysRevD.98.043526
  [arXiv:1708.01530 [astro-ph.CO]].
  %%CITATION = doi:10.1103/PhysRevD.98.043526;%%
  %407 citations counted in INSPIRE as of 19 Nov 2019


%\cite{1705.00743}
\bibitem{1705.00743} 
  Y.~Omori {\it et al.},
  %``A 2500 deg$^2$ CMB Lensing Map from Combined South Pole Telescope and Planck Data,''
  Astrophys.\ J.\  {\bf 849}, no. 2, 124 (2017)
  doi:10.3847/1538-4357/aa8d1d
  [arXiv:1705.00743 [astro-ph.CO]].
  %%CITATION = doi:10.3847/1538-4357/aa8d1d;%%
  %30 citations counted in INSPIRE as of 19 Nov 2019


%\cite{1110.3193}
\bibitem{1110.3193} 
  R.~Laureijs {\it et al.} [EUCLID Collaboration],
  %``Euclid Definition Study Report,''
  arXiv:1110.3193 [astro-ph.CO].
  %%CITATION = ARXIV:1110.3193;%%
  %1324 citations counted in INSPIRE as of 19 Nov 2019


%\cite{1611.00036}
\bibitem{1611.00036} 
  A.~Aghamousa {\it et al.} [DESI Collaboration],
  %``The DESI Experiment Part I: Science,Targeting, and Survey Design,''
  arXiv:1611.00036 [astro-ph.IM].
  %%CITATION = ARXIV:1611.00036;%%
  %418 citations counted in INSPIRE as of 19 Nov 2019


%\cite{1412.4872}
\bibitem{1412.4872} 
  O.~Dor\'e {\it et al.},
  %``Cosmology with the SPHEREX All-Sky Spectral Survey,''
  arXiv:1412.4872 [astro-ph.CO].
  %%CITATION = ARXIV:1412.4872;%%
  %134 citations counted in INSPIRE as of 19 Nov 2019


%\cite{Jain:1996st}
\bibitem{Jain:1996st} 
  B.~Jain and U.~Seljak,
  %``Cosmological model predictions for weak lensing: Linear and nonlinear regimes,''
  Astrophys.\ J.\  {\bf 484}, 560 (1997)
  doi:10.1086/304372
  [astro-ph/9611077].
  %%CITATION = doi:10.1086/304372;%%
  %263 citations counted in INSPIRE as of 19 Nov 2019


%\cite{Blas:2011rf}
\bibitem{Blas:2011rf} 
  D.~Blas, J.~Lesgourgues and T.~Tram,
  %``The Cosmic Linear Anisotropy Solving System (CLASS) II: Approximation schemes,''
  JCAP {\bf 1107}, 034 (2011)
  doi:10.1088/1475-7516/2011/07/034
  [arXiv:1104.2933 [astro-ph.CO]].
  %%CITATION = doi:10.1088/1475-7516/2011/07/034;%%
  %674 citations counted in INSPIRE as of 19 Nov 2019


%\cite{1307.1459}
\bibitem{1307.1459} 
  E.~Di Dio, F.~Montanari, J.~Lesgourgues and R.~Durrer,
  %``The CLASSgal code for Relativistic Cosmological Large Scale Structure,''
  JCAP {\bf 1311}, 044 (2013)
  doi:10.1088/1475-7516/2013/11/044
  [arXiv:1307.1459 [astro-ph.CO]].
  %%CITATION = doi:10.1088/1475-7516/2013/11/044;%%
  %99 citations counted in INSPIRE as of 19 Nov 2019


%\cite{1208.2701}
\bibitem{1208.2701} 
  R.~Takahashi, M.~Sato, T.~Nishimichi, A.~Taruya and M.~Oguri,
  %``Revising the Halofit Model for the Nonlinear Matter Power Spectrum,''
  Astrophys.\ J.\  {\bf 761}, 152 (2012)
  doi:10.1088/0004-637X/761/2/152
  [arXiv:1208.2701 [astro-ph.CO]].
  %%CITATION = doi:10.1088/0004-637X/761/2/152;%%
  %477 citations counted in INSPIRE as of 19 Nov 2019


%\cite{1807.06209}
\bibitem{1807.06209} 
  N.~Aghanim {\it et al.} [Planck Collaboration],
  %``Planck 2018 results. VI. Cosmological parameters,''
  arXiv:1807.06209 [astro-ph.CO].
  %%CITATION = ARXIV:1807.06209;%%
  %1801 citations counted in INSPIRE as of 19 Nov 2019


%\cite{Ma:2005rc}
\bibitem{Ma:2005rc} 
  Z.~M.~Ma, W.~Hu and D.~Huterer,
  %``Effect of photometric redshift uncertainties on weak lensing tomography,''
  Astrophys.\ J.\  {\bf 636}, 21 (2005)
  doi:10.1086/497068
  [astro-ph/0506614].
  %%CITATION = doi:10.1086/497068;%%
  %210 citations counted in INSPIRE as of 19 Nov 2019


%\cite{Pozzetti:2016cch}
\bibitem{Pozzetti:2016cch} 
  L.~Pozzetti {\it et al.},
  %``Modelling the number density of Hα emitters for future spectroscopic near-IR space missions,''
  Astron.\ Astrophys.\  {\bf 590}, A3 (2016)
  doi:10.1051/0004-6361/201527081
  [arXiv:1603.01453 [astro-ph.GA]].
  %%CITATION = doi:10.1051/0004-6361/201527081;%%
  %29 citations counted in INSPIRE as of 19 Nov 2019


%\cite{1903.02030}
\bibitem{1903.02030} 
  A.~Merson, A.~Smith, A.~Benson, Y.~Wang and C.~M.~Baugh,
  %``Linear bias forecasts for emission line cosmological surveys,''
  Mon.\ Not.\ Roy.\ Astron.\ Soc.\  {\bf 486}, no. 4, 5737 (2019)
  doi:10.1093/mnras/stz1204
  [arXiv:1903.02030 [astro-ph.CO]].
  %%CITATION = doi:10.1093/mnras/stz1204;%%
  %3 citations counted in INSPIRE as of 19 Nov 2019


%\cite{1206.1225}
\bibitem{1206.1225} 
  L.~Amendola {\it et al.} [Euclid Theory Working Group],
  %``Cosmology and fundamental physics with the Euclid satellite,''
  Living Rev.\ Rel.\  {\bf 16}, 6 (2013)
  doi:10.12942/lrr-2013-6
  [arXiv:1206.1225 [astro-ph.CO]].
  %%CITATION = doi:10.12942/lrr-2013-6;%%
  %572 citations counted in INSPIRE as of 19 Nov 2019


%\cite{1807.06205}
\bibitem{1807.06205} 
  Y.~Akrami {\it et al.} [Planck Collaboration],
  %``Planck 2018 results. I. Overview and the cosmological legacy of Planck,''
  arXiv:1807.06205 [astro-ph.CO].
  %%CITATION = ARXIV:1807.06205;%%
  %230 citations counted in INSPIRE as of 19 Nov 2019


%\cite{1808.07445}
\bibitem{1808.07445} 
  P.~Ade {\it et al.} [Simons Observatory Collaboration],
  %``The Simons Observatory: Science goals and forecasts,''
  JCAP {\bf 1902}, 056 (2019)
  doi:10.1088/1475-7516/2019/02/056
  [arXiv:1808.07445 [astro-ph.CO]].
  %%CITATION = doi:10.1088/1475-7516/2019/02/056;%%
  %152 citations counted in INSPIRE as of 19 Nov 2019


%\cite{1610.02743}
\bibitem{1610.02743} 
  K.~N.~Abazajian {\it et al.} [CMB-S4 Collaboration],
  %``CMB-S4 Science Book, First Edition,''
  arXiv:1610.02743 [astro-ph.CO].
  %%CITATION = ARXIV:1610.02743;%%
  %628 citations counted in INSPIRE as of 19 Nov 2019


%\cite{Matsumura:2016sri}
\bibitem{Matsumura:2016sri} 
  T.~Matsumura {\it et al.},
  %``LiteBIRD: Mission Overview and Focal Plane Layout,''
  J.\ Low.\ Temp.\ Phys.\  {\bf 184}, no. 3-4, 824 (2016).
  doi:10.1007/s10909-016-1542-8
  %%CITATION = doi:10.1007/s10909-016-1542-8;%%
  %43 citations counted in INSPIRE as of 19 Nov 2019


%\cite{1902.10541}
\bibitem{1902.10541} 
  S.~Hanany {\it et al.} [NASA PICO Collaboration],
  %``PICO: Probe of Inflation and Cosmic Origins,''
  arXiv:1902.10541 [astro-ph.IM].
  %%CITATION = ARXIV:1902.10541;%%
  %47 citations counted in INSPIRE as of 19 Nov 2019


%\cite{1909.01591}
\bibitem{1909.01591} 
  J.~Delabrouille {\it et al.},
  %``Microwave Spectro-Polarimetry of Matter and Radiation across Space and Time,''
  arXiv:1909.01591 [astro-ph.CO].
  %%CITATION = ARXIV:1909.01591;%%
  %2 citations counted in INSPIRE as of 19 Nov 2019


%\cite{astro-ph/0301031}
\bibitem{astro-ph/0301031} 
  T.~Okamoto and W.~Hu,
  %``CMB lensing reconstruction on the full sky,''
  Phys.\ Rev.\ D {\bf 67}, 083002 (2003)
  doi:10.1103/PhysRevD.67.083002
  [astro-ph/0301031].
  %%CITATION = doi:10.1103/PhysRevD.67.083002;%%
  %303 citations counted in INSPIRE as of 19 Nov 2019


%\cite{Adam:2015rua}
\bibitem{Adam:2015rua} 
  R.~Adam {\it et al.} [Planck Collaboration],
  %``Planck 2015 results. I. Overview of products and scientific results,''
  Astron.\ Astrophys.\  {\bf 594}, A1 (2016)
  doi:10.1051/0004-6361/201527101
  [arXiv:1502.01582 [astro-ph.CO]].
  %%CITATION = doi:10.1051/0004-6361/201527101;%%
  %831 citations counted in INSPIRE as of 19 Nov 2019


%\cite{1612.08270}
\bibitem{1612.08270} 
  F.~Finelli {\it et al.} [CORE Collaboration],
  %``Exploring cosmic origins with CORE: Inflation,''
  JCAP {\bf 1804}, 016 (2018)
  doi:10.1088/1475-7516/2018/04/016
  [arXiv:1612.08270 [astro-ph.CO]].
  %%CITATION = doi:10.1088/1475-7516/2018/04/016;%%
  %103 citations counted in INSPIRE as of 19 Nov 2019


%\cite{Limber:1954zz}
\bibitem{Limber:1954zz} 
  D.~N.~Limber,
  %``The Analysis of Counts of the Extragalactic Nebulae in Terms of a Fluctuating Density Field. II,''
  Astrophys.\ J.\  {\bf 119}, 655 (1954).
  doi:10.1086/145870
  %%CITATION = doi:10.1086/145870;%%
  %296 citations counted in INSPIRE as of 19 Nov 2019

%\cite{Challinor:2011bk}
\bibitem{Challinor:2011bk} 
  A.~Challinor and A.~Lewis,
  %``The linear power spectrum of observed source number counts,''
  Phys.\ Rev.\ D {\bf 84}, 043516 (2011)
  doi:10.1103/PhysRevD.84.043516
  [arXiv:1105.5292 [astro-ph.CO]].
  %%CITATION = doi:10.1103/PhysRevD.84.043516;%%
  %294 citations counted in INSPIRE as of 20 Nov 2019

%\cite{Ballardini:2018cho}
\bibitem{Ballardini:2018cho} 
  M.~Ballardini and R.~Maartens,
  %``Measuring ISW with next-generation radio surveys,''
  Mon.\ Not.\ Roy.\ Astron.\ Soc.\  {\bf 485}, 1339 (2019)
  doi:10.1093/mnras/stz480
  [arXiv:1812.01636 [astro-ph.CO]].
  %%CITATION = doi:10.1093/mnras/stz480;%%
  %6 citations counted in INSPIRE as of 20 Nov 2019

%\cite{Pourtsidou:2015qaa}
\bibitem{Pourtsidou:2015qaa} 
  A.~Pourtsidou, D.~Bacon and R.~Crittenden,
  %``Cross-correlation cosmography with intensity mapping of the neutral hydrogen 21 cm emission,''
  Phys.\ Rev.\ D {\bf 92}, no. 10, 103506 (2015)
  doi:10.1103/PhysRevD.92.103506
  [arXiv:1506.02615 [astro-ph.CO]].
  %%CITATION = doi:10.1103/PhysRevD.92.103506;%%
  %5 citations counted in INSPIRE as of 26 Nov 2019

%\cite{astro-ph/9603021}
\bibitem{astro-ph/9603021} 
  M.~Tegmark, A.~Taylor and A.~Heavens,
  %``Karhunen-Loeve eigenvalue problems in cosmology: How should we tackle large data sets?,''
  Astrophys.\ J.\  {\bf 480}, 22 (1997)
  doi:10.1086/303939
  [astro-ph/9603021].
  %%CITATION = doi:10.1086/303939;%%
  %647 citations counted in INSPIRE as of 19 Nov 2019


%\cite{Jungman:1995bz}
\bibitem{Jungman:1995bz} 
  G.~Jungman, M.~Kamionkowski, A.~Kosowsky and D.~N.~Spergel,
  %``Cosmological parameter determination with microwave background maps,''
  Phys.\ Rev.\ D {\bf 54}, 1332 (1996)
  doi:10.1103/PhysRevD.54.1332
  [astro-ph/9512139].
  %%CITATION = doi:10.1103/PhysRevD.54.1332;%%
  %450 citations counted in INSPIRE as of 19 Nov 2019


%\cite{Eisenstein:1998hr}
\bibitem{Eisenstein:1998hr} 
  D.~J.~Eisenstein, W.~Hu and M.~Tegmark,
  %``Cosmic complementarity: Joint parameter estimation from CMB experiments and redshift surveys,''
  Astrophys.\ J.\  {\bf 518}, 2 (1999)
  doi:10.1086/307261
  [astro-ph/9807130].
  %%CITATION = doi:10.1086/307261;%%
  %346 citations counted in INSPIRE as of 19 Nov 2019


%\cite{Kitching:2008eq}
\bibitem{Kitching:2008eq} 
  T.~D.~Kitching, A.~Amara, F.~B.~Abdalla, B.~Joachimi and A.~Refregier,
  %``Cosmological Systematics Beyond Nuisance Parameters : Form Filling Functions,''
  Mon.\ Not.\ Roy.\ Astron.\ Soc.\  {\bf 399}, 2107 (2009)
  doi:10.1111/j.1365-2966.2009.15408.x
  [arXiv:0812.1966 [astro-ph]].
  %%CITATION = doi:10.1111/j.1365-2966.2009.15408.x;%%
  %41 citations counted in INSPIRE as of 19 Nov 2019
  
 %\cite{Dizgah:2016bgm}
\bibitem{Dizgah:2016bgm} 
  A.~Moradinezhad Dizgah and R.~Durrer,
  %``Lensing corrections to the $E_g(z)$ statistics from large scale structure,''
  JCAP {\bf 1609}, 035 (2016)
  doi:10.1088/1475-7516/2016/09/035
  [arXiv:1604.08914 [astro-ph.CO]].
  %%CITATION = doi:10.1088/1475-7516/2016/09/035;%%
  %11 citations counted in INSPIRE as of 19 Nov 2019
  
 %\cite{Ghosh:2018ijm}
\bibitem{Ghosh:2018ijm} 
  B.~Ghosh and R.~Durrer,
  %``The observable $E_g$ statistics,''
  JCAP {\bf 1906}, 010 (2019)
  doi:10.1088/1475-7516/2019/06/010
  [arXiv:1812.09546 [astro-ph.CO]].
  %%CITATION = doi:10.1088/1475-7516/2019/06/010;%%
  %2 citations counted in INSPIRE as of 19 Nov 2019
\end{thebibliography}
\end{document}